\documentclass[journal]{IEEEtaes}
\usepackage{cite}
\usepackage{amssymb,amsmath,latexsym,amsfonts,mathtools}

\usepackage{graphicx}
\usepackage{float,subcaption}
\usepackage{textcomp}
\usepackage{xcolor}
\usepackage{hyperref}
\hypersetup{colorlinks, breaklinks, citecolor=blue, linkcolor=blue, urlcolor=blue}
\usepackage[nameinlink]{cleveref}
\Crefformat{figure}{#2Fig.~#1#3}
\Crefmultiformat{figure}{Figs.~#2#1#3}{ and~#2#1#3}{, #2#1#3}{ and~#2#1#3}
\usepackage{tikz}
\usetikzlibrary{automata, shapes, arrows, calc, arrows.meta, fit, positioning}
\usepackage{amsthm}
\theoremstyle{plain}
\newtheorem{theorem}{Theorem}
\newtheorem{lemma}{Lemma}

\newtheorem{proposition}[theorem]{Proposition}
\newtheorem*{problem*}{Problem}
\theoremstyle{remark}
\newtheorem{remark}{Remark}

\theoremstyle{definition}

\usepackage{siunitx}
\newcommand{\sign}{\mathrm{sign}}
\usepackage{mathrsfs}
\usepackage{newtxmath}
\usepackage{booktabs}
\usepackage{accents}
\newcommand\munderbar[1]{%
  \underaccent{\bar}{#1}}

\allowdisplaybreaks
\IEEEoverridecommandlockouts

\begin{document}

\title{Cooperative Nonlinear Guidance Strategies for Guaranteed Pursuit-Evasion} 

\author{Saurabh Kumar}
\affil{Indian Institute of Technology Bombay, Powai, Mumbai 400076, India} 
\author{Shashi Ranjan Kumar}
\member{Senior Member, IEEE}
\affil{Indian Institute of Technology Bombay, Powai, Mumbai 400076, India} 
\author{Abhinav Sinha}
\member{Senior Member, IEEE}
\affil{University of Cincinnati, Cincinnati, OH 45221, USA}


\corresp{{\itshape (Corresponding author: S. Kumar)}}

\authoraddress{S. Kumar and S. R. Kumar are with the Intelligent Systems and Control Lab, Department of Aerospace Engineering, Indian Institute of Technology Bombay, Powai, Mumbai 400076, India. (e-mails: \href{saurabh.k@aero.iitb.ac.in}{saurabh.k@aero.iitb.ac.in}, \href{srk@aero.iitb.ac.in}{srk@aero.iitb.ac.in}). A. Sinha is with GALACxIS Lab, Department of Aerospace Engineering and Engineering Mechanics, University of Cincinnati, OH 45221, USA. (e-mail: \href{abhinav.sinha@uc.edu}{abhinav.sinha@uc.edu}).}


\maketitle
\begin{abstract}
This paper investigates a pursuit-evasion problem involving three agents: a pursuer, an evader, and a defender. Cooperative guidance laws are developed for the evader–defender team that guarantee interception of the pursuer by the defender before it reaches the vicinity of the evader. Unlike heuristic methods, optimal control, differential game formulation, and recently proposed time-constrained guidance techniques, a geometry-based solution is proposed to safeguard the evader from the pursuer's incoming threat. The proposed strategy is computationally efficient and expected to be scalable as the number of agents increases. Another notable feature of the proposed strategy is that the evader–defender team does not require knowledge of the pursuer’s strategy, yet the pursuer's interception is guaranteed for arbitrary initial engagement geometries. It is further shown that the relevant error variables for the evader–defender team (or individual) converge to zero at a prespecified finite time that can be exactly prescribed prior to the three-body engagement. Finally, the effectiveness of the proposed cooperative pursuit-evasion strategy is demonstrated through simulations across diverse engagement scenarios.
\end{abstract}

\begin{IEEEkeywords}
Pursuit-evasion, guidance, autonomy, aerospace, multi-agent systems, aircraft defense.
\end{IEEEkeywords}

\section{Introduction}\label{sec:introduction}
The pursuit-evasion problem has long been a subject of study in game theory, robotics, aerospace, and control engineering \cite{isaacs1999differential}. While the pursuit-evasion scenarios involving only two participants (see \cite{7855582,4383577} and references therein) have been extensively studied, the extension to a three-body pursuit-evasion problem adds a layer of complexity and requires higher autonomy and strategic decision-making. Such engagements involve three agents, namely, a pursuer, an evader, and a defender. The pursuer seeks to intercept the evader, while the defender assists the evader in avoiding interception.

Early research on the kinematics of three-body engagement can be found in \cite{4101686}. In \cite{4101686}, a closed-form solution was derived for constant-bearing collision courses, while the work in \cite{4102335} focused on determining the intercept point's location in the evader-centered reference frame. For three-agent engagements, optimal control-based formulations with specific objectives, such as minimizing energy or cost, were employed in cooperative guidance strategies, as discussed in \cite{doi:10.2514/1.G001083,doi:10.2514/1.51765,doi:10.2514/1.49515,doi:10.2514/1.58531,doi:10.2514/1.61832}. In \cite{doi:10.2514/1.G001083}, the authors presented a cooperative optimal guidance strategy integrated with a differential game formulation to maximize the separation between pursuer and evader. In \cite{doi:10.2514/1.51765}, optimal cooperative pursuit-evasion strategies for the defender-evader team were proposed, considering arbitrary order-linearized dynamics for each agent. It was assumed that the pursuer's guidance strategy was known in this case. The work in \cite{doi:10.2514/1.49515} introduced a multiple-model adaptive estimator approach for cooperative information-sharing between the evader and defender to estimate the likely linear guidance strategy of the pursuer. The work in \cite{doi:10.2514/1.58531} discussed three-layer cooperation between the defender and evader and explored information exchange between them, whereas that in \cite{doi:10.2514/1.61832} provided algebraic conditions under which the pursuer could capture the evader by bypassing the defender. Note that most of these strategies relied on linearized dynamics, simplifying guidance design but potentially limiting their applicability in diverse operating conditions and scenarios with significant heading errors.

Guidance strategies developed in a nonlinear context can overcome these limitations and enhance performance, e.g., by relaxing small heading error assumptions and accounting for turn constraints. Notable works in this regard include \cite{6315051,doi:10.2514/1.G000659,9274339,doi:10.1007/s10846-022-01570-y,doi:10.1016/j.ast.2020.105787}. In \cite{6315051}, the authors introduced a sliding mode control-based terminal intercept guidance and autopilot design for defenders to protect the evader from incoming pursuers. Another nonlinear guidance strategy employing sliding mode control was discussed in \cite{doi:10.2514/1.G000659}. In \cite{9274339}, a nonlinear guidance strategy was explored for scenarios where multiple defenders simultaneously intercept the pursuer before it reaches the evader.  In \cite{doi:10.1007/s10846-022-01570-y}, nonlinear feedback laws were developed to guide the evader on a collision course with the pursuer as a decoy, allowing the defender to intercept the pursuer before it captures the evader. This approach also provided the defender with the flexibility to adopt either a defensive or aggressive stance based on mission requirements. Another nonlinear guidance strategy, based on relative line-of-sight error and time-to-go deviation between the evader and the defender, was presented in \cite{doi:10.1016/j.ast.2020.105787}. 
It is important to note that these guidance strategies rely on time-to-estimates, which may not always be available with the required precision, potentially affecting their effectiveness.

Geometrical approaches have also found application in three-agent pursuit-evasion scenarios. For instance, in \cite{doi:10.2514/6.2010-7876,9301417}, a method centered on line-of-sight guidance, a three-point guidance strategy, was explored. These three points were defined as the evader, the defender, and the pursuer. The approach demonstrated that if the defender remains aligned with the line-of-sight connecting the evader and the pursuer, the interception of the pursuer is assured before it nears the evader. In a related vein, a modified version of the command-to-line-of-sight guidance approach, incorporating optimal control theory and velocity error feedback, was introduced in \cite{doi:10.2514/1.58566}. In \cite{10643724}, a two-stage guidance scheme based on the line-of-sight method for two-on-two engagements is presented.  Furthermore, in \cite{doi:10.2514/1.G006705}, the authors presented a guidance strategy based on a barrier Lyapunov function to protect the evader from a pursuer. It is worth noting that the use of the barrier function imposed restrictions on some engagement variables, such as the defender's initial heading angle, thereby potentially limiting the target set in which the game could terminate. Motivated by these results, the focus of the current paper is on analyzing and presenting a simple and intuitive geometry-based solution for guaranteed pursuit evasion. The merits of this work can be succinctly summarized as follows:
\begin{itemize}
    \item We propose a geometrical approach to guarantee pursuit-evasion from arbitrary three-body engagement geometries. The proposed solution, which is the evader-defender cooperative guidance strategy, ensures that the defender always arrives at a certain angle within a prescribed time, regardless of the initial geometry, thereby preventing the pursuer from capturing the evader. 
    \item Unlike LOS angle-based geometric guidance, wherein the defender has to strictly maintain a fixed angle of $\pi$ with respect to the pursuer-evader LOS, the proposed strategy is less stringent and only requires a said angle to be within a broad interval of $\left[{\pi}/{2},{3\pi}/{2}\right]$. This allows the defender to have more flexibility in desired angle selection depending on engagement scenarios.
    \item We propose two levels of cooperation between the evader–defender team, which can be employed depending on the available communication between them and the required level of simplicity in implementation to safeguard the evader from the incoming pursuer.
    \item Within our problem framework, the dynamics governing the agents are inherently nonlinear and account for large heading angle errors and non-holonomic constraints. Consequently, the steering control variable for each agent is its lateral acceleration, a pragmatic choice when compared to the manipulation of heading angles. Such consideration is more practical in the context of aerial vehicles, e.g., in aircraft defense.
    \item The proposed strategy has an intuitive appeal, is expected to be computationally efficient, and sets itself apart from heuristic methodologies, optimal control strategies, and formulations rooted in differential games, where analytical solutions may cease to exist due to challenges associated with nonlinearity and the absence of knowledge about the pursuer's strategy.
    \item The proposed geometry-based solutions are versatile and can be applied to a wide range of pursuit-evasion scenarios involving different numbers of agents, dimensions, and constraints. By analyzing the geometry of the problem, the results in this paper open up new avenues to identify necessary and sufficient conditions or configurations that lead to successful evasion or capture.
\end{itemize}

Note that even if the pursuer is sufficiently close to the evader, the evader and defender will cooperatively maneuver such that the defender can intercept the pursuer. Of course, there could be a few cases where the pursuer is too close to the evader, while the defender is too far away from them. In such pathological cases, it is apparent that the defender may not be able to defend the evader. However, it is also worthwhile to note that such cases are impractical to consider from a three-body aircraft defense scenario because if the defender is too far away, then the engagement, for all practical purposes, is only for two agents.

\section{Problem Formulation}\label{sec:problem}
\begin{figure}[!ht]
    \centering
    \includegraphics[width=.95\linewidth]{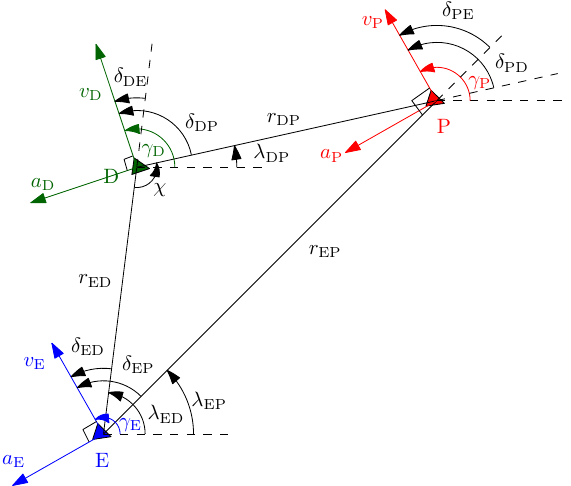}
    \caption{Geometry of a three-body pursuit–evasion engagement.}
    \label{fig:enggeo}
\end{figure}

We consider a cooperative defense problem involving nonholonomic agents, namely, a pursuer (P), an evader (E), and a defender (D). The pursuer aims to intercept the evader. In contrast, the defender's objective is to neutralize the pursuer before it reaches the vicinity of the evader. Thus, the evader and defender cooperate as allies, while the pursuer acts as the opponent. Such a scenario leads to a planar (2D) engagement, as shown in \Cref{fig:enggeo}. The agents evolve according to 
\begin{equation}\label{eq:basic}
    \dot{x}_\ell = v_\ell\cos\gamma_\ell,~\dot{y}_\ell = v_\ell\sin\gamma_\ell,~\dot{\gamma}_\ell=\dfrac{a_\ell}{v_\ell};\; \forall \ell\in\{\mathrm{P}, \mathrm{E},\mathrm{D}\},
\end{equation}
where $[x_\ell,y_\ell]^\top\in\mathbb{R}^2$, $v_\ell\in\mathbb{R}_+$, and $\gamma_\ell\in(-\pi,\pi]$ denote the position, speed, and the heading angle of the $i$\textsuperscript{th} agent, respectively, whereas $a_\ell$ is its steering control (lateral acceleration), which is assumed to be bounded. Thus, $\left|a_\ell\right|\leq a_\ell^{\max}\in\mathbb{R}_+$. This is unlike previous studies where the agents have simple dynamics, and their instantaneous heading angles were used to control them \cite{9122473,doi:10.1016/j.automatica.2011.06.010}. 

In this paper, the speeds and maneuverability of the agents are such that $v_\mathrm{P}\approx v_\mathrm{D}>v_\mathrm{E}$, and $a_\mathrm{P}^{\max}\approx a_\mathrm{D}^{\max}>a_\mathrm{E}^{\max}$, essentially implying that the pursuer and the defender are similar in capabilities with a speed advantage over the evader. As seen from \Cref{fig:enggeo}, the agents have relative separations, $r_\ell$, and the line-of-sight (LOS) angles between any two pairs of agents are $\lambda_\ell$, where $\ell=\mathrm{EP}, \mathrm{DE},$ and $\mathrm{DP}$. The engagement kinematics governing the relative motion between any two pairs can be expressed in polar coordinates as
\begin{subequations}\label{eq:engementkinematics}
\begin{align}
\dot{r}_{\mathrm{EP}} & =v_{\mathrm{P}} \cos \left(\gamma_{\mathrm{P}}-\lambda_{\mathrm{EP}}\right)-v_{\mathrm{E}} \cos \left(\gamma_{\mathrm{E}}-\lambda_{\mathrm{EP}}\right)  \nonumber  \\
& =v_{\mathrm{P}} \cos \delta_{\mathrm{PE}}-v_{\mathrm{E}} \cos \delta_{\mathrm{EP}}=v_{r_{\mathrm{EP}}},\label{eq:rdotep} \\
r_{\mathrm{EP}} \dot{\lambda}_{\mathrm{EP}} & =v_{\mathrm{P}} \sin \left(\gamma_{\mathrm{P}}-\lambda_{\mathrm{EP}}\right)-v_{\mathrm{E}} \sin \left(\gamma_{\mathrm{E}}-\lambda_{\mathrm{EP}}\right) \nonumber \\
& =v_{\mathrm{P}} \sin \delta_{\mathrm{PE}}-v_{\mathrm{E}} \sin \delta_{\mathrm{EP}}=v_{\lambda_{\mathrm{EP}}}, \label{eq:lambdadotep} \\
\dot{r}_{\mathrm{DP}} & =v_{\mathrm{P}} \cos \left(\gamma_{\mathrm{P}}-\lambda_{\mathrm{DP}}\right)-v_{\mathrm{D}} \cos \left(\gamma_{\mathrm{D}}-\lambda_{\mathrm{DP}}\right) \nonumber \\
& =v_{\mathrm{P}} \cos \delta_{\mathrm{PD}}-v_{\mathrm{D}} \cos \delta_{\mathrm{DP}}=v_{r_{\mathrm{DP}}}, \label{eq:rdotdp}\\
r_{\mathrm{DP}} \dot{\lambda}_{\mathrm{DP}} & =v_{\mathrm{P}} \sin \left(\gamma_{\mathrm{P}}-\lambda_{\mathrm{DP}}\right)-v_{\mathrm{D}} \sin \left(\gamma_{\mathrm{D}}-\lambda_{\mathrm{DP}}\right) \nonumber  \\
& =v_{\mathrm{P}} \sin \delta_{\mathrm{PD}}-v_{\mathrm{D}} \sin \delta_{\mathrm{DP}}=v_{\lambda_{\mathrm{DP}}}, \label{eq:lambdadotdp}\\
\dot{r}_{\mathrm{ED}} & =v_{\mathrm{D}} \cos \left(\gamma_{\mathrm{D}}-\lambda_{\mathrm{ED}}\right) - v_{\mathrm{E}} \cos \left(\gamma_{\mathrm{E}}-\lambda_{\mathrm{ED}}\right) \nonumber\\
& =v_{\mathrm{D}} \cos \delta_{\mathrm{DE}} - v_{\mathrm{E}} \cos \delta_{\mathrm{ED}} =v_{r_{\mathrm{ED}}}, \label{eq:rdotde} \\
r_{\mathrm{ED}} \dot{\lambda}_{\mathrm{ED}} & =v_{\mathrm{D}} \sin \left(\gamma_{\mathrm{D}}-\lambda_{\mathrm{ED}}\right) - v_{\mathrm{E}} \sin \left(\gamma_{\mathrm{E}}-\lambda_{\mathrm{ED}}\right) \nonumber  \\
& =v_{\mathrm{D}} \sin \delta_{\mathrm{DE}} - v_{\mathrm{E}} \sin \delta_{\mathrm{ED}}=v_{\lambda_{\mathrm{ED}}}, \label{eq:lambdadotde}
\end{align}
\end{subequations}
where $v_{r_{\mathrm{EP}}}, v_{r_{\mathrm{DP}}}, v_{r_{\mathrm{ED}}}, v_{\lambda_{\mathrm{EP}}}, v_{\lambda_{\mathrm{DP}}}$, and $v_{\lambda_{\mathrm{ED}}}$, represent the components of relative velocities of the relevant agents along and perpendicular to the corresponding LOS of their respective engagements. The quantities $\delta_k$ in \eqref{eq:engementkinematics} denote the corresponding lead angles and are defined as $\delta_{\mathrm{PE}}=\gamma_\mathrm{P}-\lambda_{\mathrm{EP}}$, $\delta_{\mathrm{EP}}=\gamma_\mathrm{E}-\lambda_{\mathrm{EP}}$, $\delta_{\mathrm{PD}}=\gamma_\mathrm{P}-\lambda_{\mathrm{DP}}$, $\delta_{\mathrm{DP}}=\gamma_\mathrm{D}-\lambda_{\mathrm{DP}}$, $\delta_{\mathrm{ED}}=\gamma_\mathrm{E}-\lambda_{\mathrm{ED}}$, $\delta_{\mathrm{DE}}=\gamma_\mathrm{D}-\lambda_{\mathrm{ED}}$. Note that \eqref{eq:rdotep}--\eqref{eq:lambdadotep} describe the equations of motion between the evader and the pursuer, while  \eqref{eq:rdotdp}--\eqref{eq:lambdadotdp}  represent the same for the defender-pursuer pair. On the other hand, the cooperative engagement between the evader-defender pair is described by \eqref{eq:rdotde}--\eqref{eq:lambdadotde}.

 The goal of this paper is to find a feasible nonlinear guidance strategy such that the evader-defender team could cooperatively ensure the evader's survival regardless of the pursuer's guidance law. This essentially means that we are interested in designing $a_\mathrm{E}$ and $a_\mathrm{D}$ such that a target set $\mathscr{T}=\{\mathscr{S}\,|\,r_\mathrm{DP}(t_f)=0\}$, where  $\mathcal{S}$ is the set of relevant states of the agents (e.g., position, velocity, heading, range, LOS, etc.) and $t_f$ is the time when the defender captures the pursuer, can be reached. Note that the evader-defender alliance has no knowledge of the pursuer's guidance law. However, each agent can measure the relative information of every other agent.

\section{Main Results}\label{sec:main}
\begin{figure}[t!]
    \centering
    \includegraphics[width=.65\linewidth]{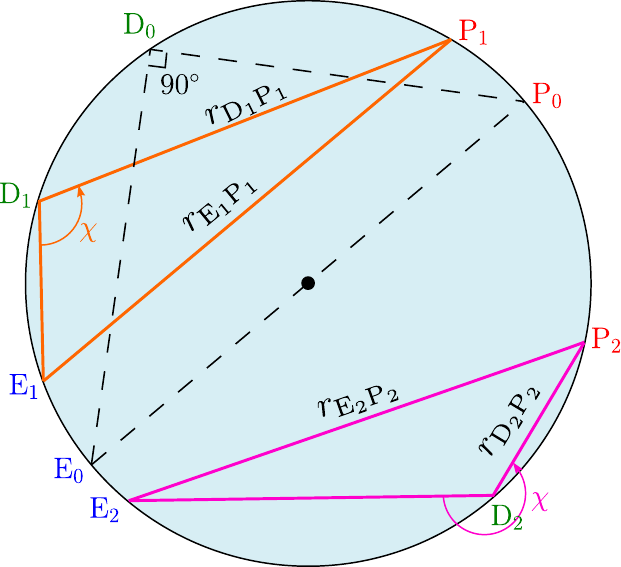}
    \caption{Illustration of the angle $\chi$.}
    \label{fig:circle}
\end{figure}

Consider the triangle formed by joining the pursuer, the evader, and the defender at any given point of time, as illustrated in $\triangle$ P\textsubscript{1}E\textsubscript{1}D\textsubscript{1} in \Cref{fig:circle}. A circle can be thought of as circumscribing this triangle such that the relative distances between the pair of agents form chords in the circle. The relative distance between the pursuer and the evader, $r_{\mathrm{E}_1\mathrm{P}_1}$, can be considered the base of $\triangle$ P\textsubscript{1}E\textsubscript{1}D\textsubscript{1}, whereas the other two relative distances represent the other sides of the triangle. It is immediate from basic geometry that when a chord divides a circle, it creates two segments-- one with an acute angle subtended (major segment) and the other with an obtuse angle (minor segment). The diameter is the longest chord, and it subtends a right angle in both segments. Following this rule, it is apparent that $\angle$ E\textsubscript{1}D\textsubscript{1}P\textsubscript{1} is obtuse, whereas $\angle$ E\textsubscript{0}D\textsubscript{0}P\textsubscript{0} in $\triangle$ P\textsubscript{0}E\textsubscript{0}D\textsubscript{0} subtends $90^\circ$ as E\textsubscript{0}P\textsubscript{0} passes through the diameter of the circle.

Let us define an angle $\chi$ such that it is the angle subtended by the evader-defender and the defender-pursuer LOS at any segment, which is measured positive in the counterclockwise sense. Referring to \Cref{fig:circle}, it is the $\angle$ E\textsubscript{1}D\textsubscript{1}P\textsubscript{1} in $\triangle$ P\textsubscript{1}E\textsubscript{1}D\textsubscript{1} and $\angle$ E\textsubscript{0}D\textsubscript{0}P\textsubscript{0}$=90^\circ$ in $\triangle$ P\textsubscript{0}E\textsubscript{0}D\textsubscript{0}. In the case of $\triangle$ P\textsubscript{2}E\textsubscript{2}D\textsubscript{2}, the fact that interior $\angle$ E\textsubscript{2}D\textsubscript{2}P\textsubscript{2} is obtuse still holds, but $\chi$ is exterior to this angle to respect the consistency of our definition. The following proposition provides a sufficient condition for the defender to capture the pursuer before the latter could intercept the evader.
\begin{proposition}\label{prop:interceptioncondition}
    Consider the three-body engagement described using \eqref{eq:engementkinematics}. If the defender maintains a positive closing speed with respect to the pursuer and attains an angle $\chi\in\left[{\pi}/{2},{3\pi}/{2}\right]$, then the pursuer's capture is guaranteed before the evader could be captured.
\end{proposition}
\begin{proof}
    Referring to \Cref{fig:circle}, it is immediate that if $\chi\in\left({\pi}/{2},{\pi}\right)$ represents a scenario depicted in  $\triangle$ P\textsubscript{1}E\textsubscript{1}D\textsubscript{1}. It follows that the side  E\textsubscript{1}P\textsubscript{1} is the longest. Hence, E\textsubscript{1}P\textsubscript{1}$>$D\textsubscript{1}P\textsubscript{1}. If the evader-defender team cooperates such that the defender maintains a fixed $\chi\in\left({\pi}/{2},{\pi}\right)$ while also ensuring a positive closing speed with respect to the pursuer ($v_{r_{\mathrm{DP}}}<0$), the $\triangle$ P\textsubscript{1}E\textsubscript{1}D\textsubscript{1} will shrink in proportion and may change its orientation, thereby generating smaller similar triangles. Eventually, D\textsubscript{1}P\textsubscript{1} will degenerate to zero (or $r_{\mathrm{D}_1\mathrm{P}_1}\to 0$) while E\textsubscript{1}P\textsubscript{1} (or $r_{\mathrm{E}_1\mathrm{P}_1}$) will still be positive. By a similar argument, it readily follows that $r_{\mathrm{D}_2\mathrm{P}_2}\to 0$ before $r_{\mathrm{E}_2\mathrm{P}_2}$ if $\chi\in\left({\pi},{3\pi}/{2}\right)$. The cases of $\chi={\pi}/{2},\pi$ are extremes but no different. When $\chi={\pi}/{2}$ (as in $\triangle$ P\textsubscript{0}E\textsubscript{0}D\textsubscript{0}), the side  E\textsubscript{0}P\textsubscript{0} is the hypotenuse and still the longest, so the same reasoning can be applied. Finally, when $\chi=\pi$, the defender is always directly between and on the pursuer-evader LOS, resulting in the pursuer's capture by the defender prior to the interception of the evader by the pursuer.
\end{proof}
From \Cref{fig:enggeo}, it is imperative to mathematically define the angle $\chi$ as 
\begin{equation}
    \chi = \pi+ \lambda_\mathrm{DP} - \lambda_\mathrm{ED}
\end{equation}
for a general case. Toward this end, our goal is to ensure that the defender always attains a fixed angle, say, $\chi^\star\in\left[\pi/2,3\pi/2\right]$. Thus, the control objective is to nullify the error variable,
\begin{equation} \label{eq:beta}
    \beta = \chi - \chi^\star = \pi+ \lambda_\mathrm{DP} - \lambda_\mathrm{ED} - \chi^\star.
\end{equation}
Additionally, we also desire that $\beta\to 0$ within a time that is independent and uniform with respect to the initial three-body engagement geometry for a guaranteed pursuit evasion.
\begin{lemma}
The dynamics of $\beta$ has a relative degree of two with respect to the steering controls of the evader and defender.
\end{lemma}
\begin{proof}
The derivative of $\beta$ can be obtained by differentiating \eqref{eq:beta} with respect to time and using the fact that $\dot{\pi} = \dot{\chi}^{\star} = 0$ as
\begin{align*}
\dot{\beta} = \dot{\lambda}_\mathrm{DP} - \dot{\lambda}_\mathrm{ED},
\end{align*}
which, on further differentiating with respect to time, yields
\begin{align}
\ddot{\beta} = \ddot{\lambda}_\mathrm{DP} - \ddot{\lambda}_\mathrm{ED}. \label{eq:ddot_beta1}
\end{align}
Next, we obtain the values of $\ddot{\lambda}_\mathrm{DP}$ and $\ddot{\lambda}_\mathrm{ED}$. The value of $\ddot{\lambda}_\mathrm{DP}$ can be obtained by differentiating the LOS rate of evader-pursuer engagement given in \eqref{eq:lambdadotdp} with respect to time as
    \begin{align}\label{eq:lambdaddotdp1}
        r_{\mathrm{DP}}\ddot{\lambda}_{\mathrm{DP}}+ \dot{r}_{\mathrm{DP}}\dot{\lambda}_{\mathrm{DP}}=v_{\mathrm{P}}
     \cos \delta_{\mathrm{PD}} \dot{\delta}_{\mathrm{PD}}-v_{\mathrm{D}}
     \cos \delta_{\mathrm{DP}} \dot{\delta}_{\mathrm{DP}}.
    \end{align}
Using the fact that $\dot{\delta}_{\mathrm{PD}}=\dot{\gamma}_{\mathrm{P}}-\dot{\lambda}_{\mathrm{DP}}$ and $\dot{\delta}_{\mathrm{DP}}=\dot{\gamma}_{\mathrm{D}}-\dot{\lambda}_{\mathrm{DP}}$, together with \eqref{eq:basic}, the expression in \eqref{eq:lambdaddotdp1} becomes
\begin{align}
r_{\mathrm{DP}}\ddot{\lambda}_{\mathrm{DP}}&=v_{\mathrm{P}}
     \cos \delta_{\mathrm{PD}}  \left( \dfrac{a_{\mathrm{ P}}}{v_\mathrm{P}}-\dot{\lambda}_{\mathrm{DP}}\right) -v_{\mathrm{D}}\cos \delta_{\mathrm{DP}}  \left( \dfrac{a_{\mathrm{ D}}}{v_\mathrm{D}}-\dot{\lambda}_{\mathrm{DP}}\right)  \nonumber\\
 &~-\dot{r}_{\mathrm{DP}}\dot{\lambda}_{\mathrm{DP}}.\label{eq:lambdaddotdp2}
\end{align}
After arranging the similar terms together in \eqref{eq:lambdaddotdp2}, we get
\begin{align}
\ddot{\lambda}_{\mathrm{DP}}=&~ \dfrac{\dot{\lambda}_{\mathrm{DP}}}{r_{\mathrm{DP}}}\left(-\dot{r}_{\mathrm{DP}}-v_{\mathrm{P}}\cos \delta_{\mathrm{PD}}+v_{\mathrm{D}}\cos \delta_{\mathrm{PD}}\right) + \dfrac{\cos \delta_{\mathrm{PD}}}{r_{\mathrm{DP}}}a_{\mathrm{P}} \nonumber \\
 &~- \dfrac{\cos \delta_{\mathrm{DP}}}{r_{\mathrm{DP}}}a_{\mathrm{D}},
 \end{align}
 which can be simplified using \eqref{eq:rdotdp} to
 \begin{align}
  \ddot{\lambda}_{\mathrm{DP}}=&~ \dfrac{-2\dot{r}_{\mathrm{DP}}\dot{\lambda}_{\mathrm{DP}}}{r_{\mathrm{DP}}}+ \left(\dfrac{\cos \delta_{\mathrm{PD}}}{r_{\mathrm{DP}}} \right)a_{\mathrm{P}}- \left(\dfrac{\cos \delta_{\mathrm{DP}}}{r_{\mathrm{DP}}}\right)a_{\mathrm{D}}. \label{eq:lambdaddotdpfinal}
 \end{align}
 By following a similar procedure, one may obtain $\ddot{\lambda}_{\mathrm{ED}}$ as 
 \begin{align}
  \ddot{\lambda}_{\mathrm{ED}}=&~ \dfrac{-2\dot{r}_{\mathrm{ED}}\dot{\lambda}_{\mathrm{ED}}}{r_{\mathrm{ED}}}- \left(\dfrac{\cos \delta_{\mathrm{ED}}}{r_{\mathrm{ED}}}\right)a_{\mathrm{E}} + \left(\dfrac{\cos \delta_{\mathrm{DE}}}{r_{\mathrm{ED}}}\right)a_{\mathrm{ D}} \label{eq:lambdaddotdefinal}.
 \end{align}
Using \eqref{eq:lambdaddotdpfinal} and \eqref{eq:lambdaddotdefinal}, the expression in \eqref{eq:ddot_beta1} becomes
\begin{align}
    \ddot{\beta}
    &=\dfrac{2\dot{r}_{\mathrm{ED}}\dot{\lambda}_{\mathrm{ED}}}{r_{\mathrm{ED}}}
 -\dfrac{2\dot{r}_{\mathrm{DP}}\dot{\lambda}_{\mathrm{DP}}}{r_{\mathrm{DP}}} + \left(\dfrac{\cos \delta_{\mathrm{ED}}}{r_{\mathrm{ED}}} \right)a_{\mathrm{E}} + \left(\dfrac{\cos\delta_{\mathrm{PD}}}{r_{\mathrm{DP}}}\right)a_{\mathrm{P}}\nonumber\\
 &~- \left( \dfrac{\cos \delta_{\mathrm{DE}}}{r_{\mathrm{ED}}}+\dfrac{\cos \delta_{\mathrm{DP}}}{r_{\mathrm{DP}}}\right)a_{\mathrm{D}},\label{eq:beta_dynamics}
\end{align}
which is affine with the steering controls of the evader and the defender, whereas that of the pursuer is an unknown quantity. 
\end{proof}

We now consider two levels of cooperation (\emph{information-level} and \emph{maneuver-level}) between the evader and the defender, depending upon the communication between them. In {information-level} cooperation, the evader independently performs maneuvers to evade the pursuer, irrespective of the defender’s actions. However, the evader shares its maneuvering information with the defender. In other words, in {information-level} cooperation, the evader is indirectly cooperating with the defender when the former does not actively participate with the latter to attain the said angle $\chi^{\star} \in \left[{\pi}/{2},{3\pi}/{2}\right]$. In {maneuver-level} cooperation, the evader actively participates (maneuvers) with the defender to attain the said angle $\chi^{\star} \in \left[{\pi}/{2},{3\pi}/{2}\right]$. We first discuss the {maneuver-level} cooperation scenario between the evader and the defender, wherein the joint cooperative efforts of the evader-defender alliance are designed.

One can write the dynamics of $\beta$ given in \eqref{eq:beta_dynamics} as
\begin{align}
    \ddot{\beta} =&~\dfrac{2\dot{r}_{\mathrm{ED}}\dot{\lambda}_{\mathrm{ED}}}{r_{\mathrm{ED}}} -\dfrac{2\dot{r}_{\mathrm{DP}}\dot{\lambda}_{\mathrm{DP}}}{r_{\mathrm{DP}}}
 + \dfrac{\cos\delta_{\mathrm{PD}}}{r_{\mathrm{DP}}}a_{\mathrm{P}}+ \mathcal{U},\label{eq:ddotchifinal}
\end{align}
where  $\mathcal{U}$ is defined as the  joint cooperative maneuver (or net control effort) given by
\begin{align}
    \mathcal{U} = \dfrac{\cos \delta_{\mathrm{ED}}}{r_{\mathrm{ED}}}a_\mathrm{E} - \left(\dfrac{\cos \delta_{\mathrm{DE}}}{r_{\mathrm{ED}}}+\dfrac{\cos \delta_{\mathrm{DP}}}{r_{\mathrm{DP}}}\right)a_\mathrm{D}.\label{eq:U}
     \end{align}
In order to nullify $\beta$, we need to design $a_{\mathrm{D}}$ and $a_{\mathrm{E}}$ such that the evader-defender team maneuvers to place the defender on a fixed $\chi^\star$ in the interval $\left[\pi/2,3\pi/2\right]$ regardless of the initial three-body engagement geometry.

Toward this objective, we consider a dual-layer sliding manifold, with the inner layer as
\begin{align}\label{eq:st}
    \munderbar{\mathcal{S}}(t)=\dot\beta(t)-g(t)
\end{align}
where $\dot{\beta}(t)=\dot{\lambda}_{\mathrm{DP}}-\dot{\lambda}_{\mathrm{DE}}$ and $g(t)$ is given by
\begin{align}\label{eq:gt}
    g(t)=\begin{cases}
        -\dfrac{k_1\,\beta(t)}{t^\star-t};& 0\leq t<t^\star,\\
        \phantom{t-t}0;& t\geq t^\star,
    \end{cases}
\end{align}
for some $k_1\in\mathbb{N}$. Note that the variable $t^{\star}$ denotes the time instant at which the error variable $\beta(t)$ nullifies to zero. As a consequence,  the defender attains a predefined angle $\chi^{\star} \in \left[\pi/2,3\pi/2\right]$ within the same time $t^{\star}$. Thereafter, to eliminate the reaching phase and achieve a global exact-time convergence, we consider the outer-layer sliding manifold as
\begin{align}
    \bar{\mathcal{S}}(t)=\munderbar{\mathcal{S}}(t)+h(t),\label{eq:sit}
\end{align}
where $h(t)$ is an auxiliary function defined as
\begin{align}\label{eq:ht}
    h(t)=\begin{cases}
        \dfrac{h(0)}{t_1^{k_2}}\left(t_1-t\right)^{k_2};& 0\leq t <t_1,\\
        \phantom{t-t}0; & t\geq t_1,
    \end{cases}
\end{align}
 for some {$k_2 \in \mathbb{N}$}, and $h(0)$ is designed to ensure $\bar{\mathcal{S}}(0)=0$, which can be obtained from \eqref{eq:sit} as $h(0)=-\munderbar{\mathcal{S}}(0)$. In \eqref{eq:ht}, $t_{1}$ is the exact time instant from which onward the inner layer sliding manifolds become identically equal to zero. In other words, the sliding mode is enforced on the inner-layer sliding manifold at time $t=t_{1}$. The fundamental concept behind eliminating the reaching phase is to ensure global exact-time convergence for the sliding variables and achieve a complete sliding mode response. This, in consequence, bolsters robustness, as sliding mode controllers typically experience diminished robustness during their reaching phase. The joint cooperative maneuver of the defender-evader team is presented in the next theorem.
\begin{theorem}[Maneuver-level Cooperation]\label{thm:joint}
 Consider the three-body engagement described using \eqref{eq:engementkinematics}, the dynamics of the error variable, \eqref{eq:ddotchifinal}, and the dual-layer sliding manifold in \eqref{eq:st} and \eqref{eq:sit}. If the defender-evader team uses \emph{maneuver-level} cooperation, and the joint cooperative maneuver of the evader-defender team is designed as
     \begin{align}
         \mathcal{U} = \dfrac{2\dot{r}_{\mathrm{DP}}\dot{\lambda}_{\mathrm{DP}}}{r_{\mathrm{DP}}} -\dfrac{2\dot{r}_{\mathrm{ED}}\dot{\lambda}_{\mathrm{ED}}}{r_{\mathrm{ED}}}+\dot{g}(t)-\dot{h}(t)-\mathcal{K} \sign(\bar{\mathcal{S}}),\label{eq:Uexp}
     \end{align}
     for some $\mathcal{K}>\sup_{t\geq 0}\left({a_{\mathrm{P}}^{\max}}/{r_{\mathrm{DP}}}\right)$, then the evader-defender team cooperatively maneuvers such that the defender converges to the desired angle $\chi^\star\in\left[{\pi}/{2},{3\pi}/{2}\right]$, within a time $t^\star$ prescribed prior to the engagement, regardless of the three-body initial engagement geometry. 
 \end{theorem}
 \begin{proof} 
 Consider a continuous, radially unbounded Lyapunov function candidate $\mathcal{V}_{1} = \frac{1}{2}\bar{\mathcal{S}}^2$. Hereafter, we streamline the notation again by dropping the arguments of variables denoting their time dependency. Upon differentiating $\mathcal{V}_{1}$ with respect to time and using the relations in \eqref{eq:st} and \eqref{eq:sit}, one may obtain
    \begin{equation*}
        \dot{\mathcal{V}}_{1} = \bar{\mathcal{S}}\dot{\bar{\mathcal{S}}} = \bar{\mathcal{S}}\left(\dot{\munderbar{\mathcal{S}}}+\dot{h}\right) = \bar{\mathcal{S}}\left(\ddot\beta-\dot{g}+\dot{h}\right),
    \end{equation*}
    which can be further written using the expression in \eqref{eq:ddotchifinal} and \eqref{eq:U} as
    \begin{align}
        \dot{\mathcal{V}}_{1} &=\bar{\mathcal{S}}\left[\dfrac{2\dot{r}_{\mathrm{ED}}\dot{\lambda}_{\mathrm{ED}}}{r_{\mathrm{ED}}} -\dfrac{2\dot{r}_{\mathrm{DP}}\dot{\lambda}_{\mathrm{DP}}}{r_{\mathrm{DP}}}
 + \dfrac{\cos\delta_{\mathrm{PD}}}{r_{\mathrm{DP}}}a_{\mathrm{P}}-\dot{g}+\dot{h}\right.\nonumber\\
 &\left.+~ \dfrac{\cos \delta_{\mathrm{ED}}}{r_{\mathrm{ED}}}a_\mathrm{E} - \left(\dfrac{\cos \delta_{\mathrm{DE}}}{r_{\mathrm{ED}}}+\dfrac{\cos \delta_{\mathrm{DP}}}{r_{\mathrm{DP}}}\right)a_\mathrm{D}\right]\nonumber\\
 \dot{\mathcal{V}}_{1} &=\bar{\mathcal{S}}\left[\dfrac{2\dot{r}_{\mathrm{ED}}\dot{\lambda}_{\mathrm{ED}}}{r_{\mathrm{ED}}} -\dfrac{2\dot{r}_{\mathrm{DP}}\dot{\lambda}_{\mathrm{DP}}}{r_{\mathrm{DP}}}
 + \dfrac{\cos\delta_{\mathrm{PD}}}{r_{\mathrm{DP}}}a_{\mathrm{P}}-\dot{g}+\dot{h} + \mathcal{U}\right] \label{eq:v1_dot_1}
    \end{align}
    Upon substituting the expression of $\mathcal{U}$ from \eqref{eq:Uexp} into \eqref{eq:v1_dot_1}, the derivative of the Lyapunov function candidate reduces to
    \begin{align}
        \dot{\mathcal{V}}_{1} =&~\bar{\mathcal{S}}\left[ -\mathcal{K}\sign(\bar{\mathcal{S}})+\dfrac{\cos\delta_{\mathrm{PD}}}{r_{\mathrm{DP}}}a_{\mathrm{P}}\right]\nonumber\\
        \leq&-\left( \mathcal{K}-\dfrac{1}{r_{\mathrm{DP}}}a_{\mathrm{P}}^{\max} \right)\lvert \bar{\mathcal{S}}\rvert~\leq 0,
    \end{align}
    since $\mathcal{K}>\sup_{t\geq 0}\dfrac{a_{\mathrm{P}}^{\max}}{r_{\mathrm{DP}}}$. This further implies that if $\bar{\mathcal{S}}(0)=0$, under the joint cooperative maneuver of the evader-defender team, $\eqref{eq:U}$, then the system's states will remain on the sliding manifold $\bar{\mathcal{S}}(t)=0~\forall~t \geq 0$.

Now, for a time instant $t=t_1<t^\star$, the reduced order dynamics from \eqref{eq:sit} can be expressed as
\begin{align}\label{eq:betadotreduced}
    \dot \beta-g=0 \implies \dot{\beta}+\dfrac{k_1\,\beta}{t^\star-t}=0.
\end{align}
On integrating \eqref{eq:betadotreduced} within suitable limits, one may obtain
\begin{align}
    \int_{\beta(0)}^{\beta(t)}\dfrac{\dot\beta}{\beta}\mathrm{d}t=-\int_0^t\dfrac{k_1}{t^\star-t}\mathrm{d}t,
\end{align}
which results in
\begin{align}\label{eq:betasolution1}
    \ln\left[\dfrac{\beta(t)}{\beta(0)}\right]=\ln\left[\dfrac{t^\star-t}{t^\star}\right]^{k_1}.
\end{align}
After some simplifications, one can express
\begin{align}\label{eq:betasolutionfinal}
    \beta(t)=\dfrac{\beta(0)}{t^{\star^{k_1}}}\left(t^\star-t\right)^{k_1},
\end{align}
which clearly shows that at time instant $t=t^\star$, $\beta(t)=0$ regardless of $\beta(0)$. Thus, the defender attains $\chi^\star$ within $t^\star$ due to the joint cooperative maneuver of the evader-defender team. This concludes the proof. 
\end{proof}
Note that $\mathcal{U}$ in \Cref{thm:joint} is the net control authority that the evader-defender pair has to maintain in order to achieve a successful interception of the pursuer. Next, we endeavor to allocate this net control effort to the evader and the defender. Needless to say, there can be several choices for the allocation of the total control effort between the defender and the evader. However, we focus on a weighted control allocation of the net control effort between the evader and the defender, subjected to the affine constraint \eqref{eq:U}. Towards that, our aim is to minimize the cost function $ \mathcal{C}$, which is defined as
\begin{align}\label{eq:costfunction}
    \mathcal{C}:= \sqrt{\left(\dfrac{a_\mathrm{E}}{\Sigma_{\mathrm{E}}}\right)^2+\left(\dfrac{a_\mathrm{D}}{\Sigma_{\mathrm{D}}}\right)^2},
\end{align}
where $\Sigma_\ell > 0$ for $\ell\in\{\mathrm{E},\mathrm{P}\}$ are weights, which are used to allocate the net control effort between the evader and the defender. For a choice $\Sigma_{\mathrm{E}}=\Sigma_{\mathrm{D}}=1$, the minimization of $ \mathcal{C}$ is equivalent to instantaneously minimizing the $\mathcal{L}_2$ norm of the net control effort. The essence of this effort allocation is presented in the next theorem.
\begin{theorem}\label{thm:allocate}
The evader-defender team cooperatively ensures the interception of the pursuer with minimum values of the lateral accelerations as 
\begin{align}
a_\mathrm{E}&=\frac{r_{\mathrm{ED}}r_{\mathrm{DP}}\cos\delta_{\mathrm{ED}}2\dot{r}_{\mathrm{DP}}\dot{\lambda}_{\mathrm{DP}}}{\Sigma^2 r_{\mathrm{DP}}^2 \cos^2\delta_{\mathrm{ED}} + \left(r_{\mathrm{DP}}\cos\delta_{\mathrm{DE}} + r_{\mathrm{ED}}\cos\delta_{\mathrm{DP}}\right)^2} \nonumber\\
&-\frac{r_{\mathrm{DP}}^2\cos\delta_{\mathrm{ED}}2\dot{r}_{\mathrm{ED}}\dot{\lambda}_{\mathrm{ED}}}{\Sigma^2 r_{\mathrm{DP}}^2 \cos^2\delta_{\mathrm{ED}} + \left(r_{\mathrm{DP}}\cos\delta_{\mathrm{DE}} + r_{\mathrm{ED}}\cos\delta_{\mathrm{DP}}\right)^2} \nonumber\\
&+\frac{r_{\mathrm{ED}}r_{\mathrm{DP}}^2\cos\delta_{\mathrm{ED}} \left( \dot{g}(t)-\dot{h}(t)-\mathcal{K} \sign(\bar{\mathcal{S}} \right)}{\Sigma^2 r_{\mathrm{DP}}^2 \cos^2\delta_{\mathrm{ED}} + \left(r_{\mathrm{DP}}\cos\delta_{\mathrm{DE}} + r_{\mathrm{ED}}\cos\delta_{\mathrm{DP}}\right)^2}, \label{eq:aE}\\
a_\mathrm{D}& =-\dfrac{\left(r_{\mathrm{DP}}\cos\delta_{\mathrm{DE}} + r_{\mathrm{ED}}\cos\delta_{\mathrm{DP}}\right)\left(r_{\mathrm{ED}} 2\dot{r}_{\mathrm{DP}}\dot{\lambda}_{\mathrm{DP}}\right)}{\left( r_{\mathrm{DP}} \cos\delta_{\mathrm{ED}} \Sigma \right)^2 + \left(r_{\mathrm{DP}}\cos\delta_{\mathrm{DE}} + r_{\mathrm{ED}}\cos\delta_{\mathrm{DP}}\right)^2} \nonumber\\
&+\dfrac{\left(r_{\mathrm{DP}}\cos\delta_{\mathrm{DE}} + r_{\mathrm{ED}}\cos\delta_{\mathrm{DP}}\right)\left(r_{\mathrm{DP}} 2\dot{r}_{\mathrm{ED}}\dot{\lambda}_{\mathrm{ED}}\right)}{\left( r_{\mathrm{DP}} \cos\delta_{\mathrm{ED}} \Sigma \right)^2 + \left(r_{\mathrm{DP}}\cos\delta_{\mathrm{DE}} + r_{\mathrm{ED}}\cos\delta_{\mathrm{DP}}\right)^2} \nonumber\\
&-\dfrac{\left(r_{\mathrm{DP}}\cos\delta_{\mathrm{DE}} + r_{\mathrm{ED}}\cos\delta_{\mathrm{DP}}\right)r_{\mathrm{DP}}r_{\mathrm{ED}}\left( \dot{g}(t)-\dot{h}(t)\right)}{\left( r_{\mathrm{DP}} \cos\delta_{\mathrm{ED}} \Sigma \right)^2 + \left(r_{\mathrm{DP}}\cos\delta_{\mathrm{DE}} + r_{\mathrm{ED}}\cos\delta_{\mathrm{DP}}\right)^2} \nonumber\\
&+\dfrac{\left(r_{\mathrm{DP}}\cos\delta_{\mathrm{DE}} + r_{\mathrm{ED}}\cos\delta_{\mathrm{DP}}\right)\left(r_{\mathrm{ED}} r_{\mathrm{DP}} \mathcal{K} \sign(\bar{\mathcal{S}})\right)}{\left( r_{\mathrm{DP}} \cos\delta_{\mathrm{ED}} \Sigma \right)^2 + \left(r_{\mathrm{DP}}\cos\delta_{\mathrm{DE}} + r_{\mathrm{ED}}\cos\delta_{\mathrm{DP}}\right)^2}. \label{eq:aD}
\end{align}
\end{theorem}
\begin{proof}
From \eqref{eq:U}, one may write
\begin{align}
\mathcal{U} = \mathfrak{B}_{1}a_\mathrm{E} - \mathfrak{B}_{2} a_\mathrm{D} \implies  a_\mathrm{E} = \dfrac{\mathcal{U} +\mathfrak{B}_{2} a_\mathrm{D}}{\mathfrak{B}_{1}}, \label{eq:aEalternate}
\end{align}
where $\mathfrak{B}_{1} = \dfrac{\cos \delta_{\mathrm{ED}}}{r_{\mathrm{ED}}}$ and $\mathfrak{B}_{2} = \left(\dfrac{\cos \delta_{\mathrm{DE}}}{r_{\mathrm{ED}}}+\dfrac{\cos \delta_{\mathrm{DP}}}{r_{\mathrm{DP}}}\right)$. Now substituting the value of $a_\mathrm{E}$ from \eqref{eq:aEalternate} in the cost function \eqref{eq:costfunction}, one may obtain
\begin{align}
\mathcal{C} = \sqrt{ \left(\dfrac{\mathcal{U} +\mathfrak{B}_{2} a_\mathrm{D}}{\mathfrak{B}_{1} \Sigma_{\mathrm{E}}} \right)^2+\left(\dfrac{a_\mathrm{D}}{\Sigma_{\mathrm{D}}}\right)^2}, \label{eq:c1}
\end{align}
Computing the partial derivative of $\mathcal{C}$ with respect to $a_\mathrm{D}$ and equating it to zero leads to
\begin{align*}
&\dfrac{ \partial\mathcal{C}}{ \partial a_{\mathrm{D}}} =\frac{1}{2\mathcal{C}}
\left[2\left(\frac{\mathcal{U}+\mathfrak{B}_{2} a_{\mathrm{D}}}{\mathfrak{B}_{1}\Sigma_{\mathrm{E}}}\right)
\frac{\mathfrak{B}_{2}}{\mathfrak{B}_{1}\Sigma_{\mathrm{E}}} + 2\left(\frac{a_{\mathrm{D}}}{\Sigma_{\mathrm{D}}}\right) \frac{1}{\Sigma_{\mathrm{D}}}\right] =0 \\
&\implies \frac{\dfrac{\mathfrak{B}_{2}\left(\mathcal{U}+\mathfrak{B}_{2}a_{\mathrm{D}}\right)}{\mathfrak{B}_{1}^{2}\Sigma_{\mathrm{E}}^{2}} +\dfrac{a_{\mathrm{D}}}{\Sigma_{\mathrm{D}}^{2}}}{\sqrt{\left(\dfrac{\mathcal{U}+\mathfrak{B}_{2}a_{\mathrm{D}}}{\mathfrak{B}_{1}\Sigma_{\mathrm{E}}}\right)^{2} +\left(\dfrac{a_{\mathrm{D}}}{\Sigma_{\mathrm{D}}}\right)^{2}}}=0\\
&\implies  \mathfrak{B}_{2}^{2} \Sigma_{\mathrm{D}}^2 a_\mathrm{D} + \mathfrak{B}_{1}^2 \Sigma_{\mathrm{E}}^2 a_\mathrm{D} = -\mathfrak{B}_{2} \Sigma_{\mathrm{D}}^2 \mathcal{U} \\
&\implies a_\mathrm{D} = \dfrac{-\mathfrak{B}_{2} \Sigma_{\mathrm{D}}^2 \mathcal{U}}{\mathfrak{B}_{1}^2 \Sigma_{\mathrm{E}}^2+ \mathfrak{B}_{2}^{2}\Sigma_{\mathrm{D}}^2} = -\dfrac{\mathfrak{B}_{2}  \mathcal{U}}{\mathfrak{B}_{1}^2 \Sigma^2+ \mathfrak{B}_{2}^{2}},
    \end{align*}
which, on substituting the value of $\mathcal{U}$ from \eqref{eq:Uexp} and $\mathfrak{B}_{1}$ and $\mathfrak{B}_{2}$ leads to the expression of the defender's lateral acceleration in \eqref{eq:aD}. To show that the value obtained is indeed corresponding to minima, we calculate the second derivative of $\mathcal{C}$ with respect to $a_{\mathrm{D}}$ as
\begin{align*}
&\frac{\partial^{2}\mathcal{C}}{\partial a_{\mathrm{D}}^{2}}
=
\frac{
\left(
\dfrac{\mathfrak{B}_{2}^{2}}{\mathfrak{B}_{1}^{2}\Sigma_{\mathrm{E}}^{2}}
+
\dfrac{1}{\Sigma_{\mathrm{D}}^{2}}
\right)
\sqrt{
\left(\dfrac{\mathcal{U}+\mathfrak{B}_{2} a_{\mathrm{D}}}{\mathfrak{B}_{1}\Sigma_{\mathrm{E}}}\right)^{2}
+
\left(\dfrac{a_{\mathrm{D}}}{\Sigma_{\mathrm{D}}}\right)^{2}
}
}{
\left[
\left(\dfrac{\mathcal{U}+\mathfrak{B}_{2} a_{\mathrm{D}}}{\mathfrak{B}_{1}\Sigma_{\mathrm{E}}}\right)^{2}
+
\left(\dfrac{a_{\mathrm{D}}}{\Sigma_{\mathrm{D}}}\right)^{2}
\right]
}
\\
&\quad
-
\frac{
\left(
\dfrac{\mathfrak{B}_{2}(\mathcal{U}+\mathfrak{B}_{2} a_{\mathrm{D}})}{\mathfrak{B}_{1}^{2}\Sigma_{\mathrm{E}}^{2}}
+
\dfrac{a_{\mathrm{D}}}{\Sigma_{\mathrm{D}}^{2}}
\right)
\left(
\dfrac{\mathfrak{B}_{2}(\mathcal{U}+\mathfrak{B}_{2} a_{\mathrm{D}})}{\mathfrak{B}_{1}^{2}\Sigma_{\mathrm{E}}^{2}}
+
\dfrac{a_{\mathrm{D}}}{\Sigma_{\mathrm{D}}^{2}}
\right)
}{
\sqrt{
\left(\dfrac{\mathcal{U}+\mathfrak{B}_{2} a_{\mathrm{D}}}{\mathfrak{B}_{1}\Sigma_{\mathrm{E}}}\right)^{2}
+
\left(\dfrac{a_{\mathrm{D}}}{\Sigma_{\mathrm{D}}}\right)^{2}
} \left[
\left(\dfrac{\mathcal{U}+\mathfrak{B}_{2} a_{\mathrm{D}}}{\mathfrak{B}_{1}\Sigma_{\mathrm{E}}}\right)^{2}
+
\left(\dfrac{a_{\mathrm{D}}}{\Sigma_{\mathrm{D}}}\right)^{2}
\right]}\\
=&\frac{
\dfrac{\mathfrak{B}_{2}^{2}}{\mathfrak{B}_{1}^{2}\Sigma_{\mathrm{E}}^{2}}
+
\dfrac{1}{\Sigma_{\mathrm{D}}^{2}}
}{
\sqrt{
\left(\dfrac{\mathcal{U}+\mathfrak{B}_{2} a_{\mathrm{D}}}{\mathfrak{B}_{1}\Sigma_{\mathrm{E}}}\right)^{2}
+
\left(\dfrac{a_{\mathrm{D}}}{\Sigma_{\mathrm{D}}}\right)^{2}
}
}
-
\frac{
\left(
\dfrac{\mathfrak{B}_{2}(\mathcal{U}+\mathfrak{B}_{2} a_{\mathrm{D}})}{\mathfrak{B}_{1}^{2}\Sigma_{\mathrm{E}}^{2}}
+
\dfrac{a_{\mathrm{D}}}{\Sigma_{\mathrm{D}}^{2}}
\right)^{2}
}{
\left[
\left(\dfrac{\mathcal{U}+\mathfrak{B}_{2} a_{\mathrm{D}}}{\mathfrak{B}_{1}\Sigma_{\mathrm{E}}}\right)^{2}
+
\left(\dfrac{a_{\mathrm{D}}}{\Sigma_{\mathrm{D}}}\right)^{2}
\right]^{3/2}
}.
\end{align*}
which can be simplified as
\begin{align*}
&\dfrac{\partial^{2}\mathcal{C}}{\partial a_{\mathrm{D}}^{2}}
=\dfrac{
\left(\dfrac{\mathfrak{B}_{2}^{2}}{\mathfrak{B}_{1}^{2}\Sigma_{\mathrm{E}}^{2}}+\dfrac{1}{\Sigma_{\mathrm{D}}^{2}}\right)
\left[\left(\dfrac{\mathcal{U}+\mathfrak{B}_{2}a_{\mathrm{D}}}{\mathfrak{B}_{1}\Sigma_{\mathrm{E}}}\right)^{2}
+\left(\dfrac{a_{\mathrm{D}}}{\Sigma_{\mathrm{D}}}\right)^{2}\right]
}{\left[
\left(\dfrac{\mathcal{U}+\mathfrak{B}_{2} a_{\mathrm{D}}}{\mathfrak{B}_{1}\Sigma_{\mathrm{E}}}\right)^{2}
+
\left(\dfrac{a_{\mathrm{D}}}{\Sigma_{\mathrm{D}}}\right)^{2}
\right]^{3/2}}\\
&~-\dfrac{\left(
\dfrac{\mathfrak{B}_{2}\left(\mathcal{U}+\mathfrak{B}_{2}a_{\mathrm{D}}\right)}{\mathfrak{B}_{1}^{2}\Sigma_{\mathrm{E}}^{2}}
+\dfrac{a_{\mathrm{D}}}{\Sigma_{\mathrm{D}}^{2}}
\right)^{2}}{\left[
\left(\dfrac{\mathcal{U}+\mathfrak{B}_{2} a_{\mathrm{D}}}{\mathfrak{B}_{1}\Sigma_{\mathrm{E}}}\right)^{2}
+
\left(\dfrac{a_{\mathrm{D}}}{\Sigma_{\mathrm{D}}}\right)^{2}
\right]^{3/2}} \\
&~=\dfrac{
\dfrac{\mathfrak{B}_{2}^{2}\left(\mathcal{U}+\mathfrak{B}_{2}a_{\mathrm{D}}\right)^{2}}{\mathfrak{B}_{1}^{4}\Sigma_{\mathrm{E}}^{4}}
+\dfrac{\mathfrak{B}_{2}^{2}a_{\mathrm{D}}^{2}}{\mathfrak{B}_{1}^{2}\Sigma_{\mathrm{E}}^{2}\Sigma_{\mathrm{D}}^{2}}
+\dfrac{\left(\mathcal{U}+\mathfrak{B}_{2}a_{\mathrm{D}}\right)^{2}}{\mathfrak{B}_{1}^{2}\Sigma_{\mathrm{E}}^{2}\Sigma_{\mathrm{D}}^{2}}
+\dfrac{a_{\mathrm{D}}^{2}}{\Sigma_{\mathrm{D}}^{4}}
}{\left[
\left(\dfrac{\mathcal{U}+\mathfrak{B}_{2} a_{\mathrm{D}}}{\mathfrak{B}_{1}\Sigma_{\mathrm{E}}}\right)^{2}
+
\left(\dfrac{a_{\mathrm{D}}}{\Sigma_{\mathrm{D}}}\right)^{2}
\right]^{3/2}} \\
&~-\dfrac{\left[
\dfrac{\mathfrak{B}_{2}^{2}\left(\mathcal{U}+\mathfrak{B}_{2}a_{\mathrm{D}}\right)^{2}}{\mathfrak{B}_{1}^{4}\Sigma_{\mathrm{E}}^{4}}
+\dfrac{2\mathfrak{B}_{2}\left(\mathcal{U}+\mathfrak{B}_{2}a_{\mathrm{D}}\right)a_{\mathrm{D}}}{\mathfrak{B}_{1}^{2}\Sigma_{\mathrm{E}}^{2}\Sigma_{\mathrm{D}}^{2}}
+\dfrac{a_{\mathrm{D}}^{2}}{\Sigma_{\mathrm{D}}^{4}}
\right]}{\left[
\left(\dfrac{\mathcal{U}+\mathfrak{B}_{2} a_{\mathrm{D}}}{\mathfrak{B}_{1}\Sigma_{\mathrm{E}}}\right)^{2}
+
\left(\dfrac{a_{\mathrm{D}}}{\Sigma_{\mathrm{D}}}\right)^{2}
\right]^{3/2}}\\
&=\dfrac{
\dfrac{\mathfrak{B}_{2}^{2}\left(\mathcal{U}+\mathfrak{B}_{2}a_{\mathrm{D}}\right)^{2}}{\mathfrak{B}_{1}^{4}\Sigma_{\mathrm{E}}^{4}}
+\dfrac{\mathfrak{B}_{2}^{2}a_{\mathrm{D}}^{2}}{\mathfrak{B}_{1}^{2}\Sigma_{\mathrm{E}}^{2}\Sigma_{\mathrm{D}}^{2}}
+\dfrac{\left(\mathcal{U}+\mathfrak{B}_{2}a_{\mathrm{D}}\right)^{2}}{\mathfrak{B}_{1}^{2}\Sigma_{\mathrm{E}}^{2}\Sigma_{\mathrm{D}}^{2}}
+\dfrac{a_{\mathrm{D}}^{2}}{\Sigma_{\mathrm{D}}^{4}}
}{\left[
\left(\dfrac{\mathcal{U}+\mathfrak{B}_{2} a_{\mathrm{D}}}{\mathfrak{B}_{1}\Sigma_{\mathrm{E}}}\right)^{2}
+
\left(\dfrac{a_{\mathrm{D}}}{\Sigma_{\mathrm{D}}}\right)^{2}
\right]^{3/2}}\\
&-\dfrac{\left[
\dfrac{\mathfrak{B}_{2}^{2}\left(\mathcal{U}+\mathfrak{B}_{2}a_{\mathrm{D}}\right)^{2}}{\mathfrak{B}_{1}^{4}\Sigma_{\mathrm{E}}^{4}}
+\dfrac{2\mathfrak{B}_{2}\left(\mathcal{U}+\mathfrak{B}_{2}a_{\mathrm{D}}\right)a_{\mathrm{D}}}{\mathfrak{B}_{1}^{2}\Sigma_{\mathrm{E}}^{2}\Sigma_{\mathrm{D}}^{2}}
+\dfrac{a_{\mathrm{D}}^{2}}{\Sigma_{\mathrm{D}}^{4}}
\right]}{\left[
\left(\dfrac{\mathcal{U}+\mathfrak{B}_{2} a_{\mathrm{D}}}{\mathfrak{B}_{1}\Sigma_{\mathrm{E}}}\right)^{2}
+
\left(\dfrac{a_{\mathrm{D}}}{\Sigma_{\mathrm{D}}}\right)^{2}
\right]^{3/2}},
\end{align*}
which, on further simplification, results in
\begin{align}
\dfrac{\partial^{2}\mathcal{C}}{\partial a_{\mathrm{D}}^{2}}&=\dfrac{
\mathfrak{B}_{2}^{2}a_{\mathrm{D}}^{2}
+\left(\mathcal{U}+\mathfrak{B}_{2}a_{\mathrm{D}}\right)^{2}
-2\mathfrak{B}_{2}\left(\mathcal{U}+\mathfrak{B}_{2}a_{\mathrm{D}}\right)a_{\mathrm{D}}
}{\mathfrak{B}_{1}^{2}\Sigma_{\mathrm{E}}^{2}\Sigma_{\mathrm{D}}^{2}\left[
\left(\dfrac{\mathcal{U}+\mathfrak{B}_{2} a_{\mathrm{D}}}{\mathfrak{B}_{1}\Sigma_{\mathrm{E}}}\right)^{2}
+
\left(\dfrac{a_{\mathrm{D}}}{\Sigma_{\mathrm{D}}}\right)^{2}
\right]^{3/2}},\nonumber \\
&=\dfrac{\mathcal{U}^{2}}{\mathfrak{B}_{1}^{2}\Sigma_{\mathrm{E}}^{2}\Sigma_{\mathrm{D}}^{2}\left[
\left(\dfrac{\mathcal{U}+\mathfrak{B}_{2} a_{\mathrm{D}}}{\mathfrak{B}_{1}\Sigma_{\mathrm{E}}}\right)^{2}
+
\left(\dfrac{a_{\mathrm{D}}}{\Sigma_{\mathrm{D}}}\right)^{2}
\right]^{3/2}}. \label{eq:second_partial_ad}
\end{align}
It can be readily verified from \eqref{eq:second_partial_ad} that the second derivative is positive for all $\mathcal{U}\neq 0$.  Thus, the value obtained is indeed a minimum. Thereafter, one may use \eqref{eq:aEalternate} to arrive at \eqref{eq:aE}, and by following a similar procedure, one can obtain the value of $a_{\mathrm{E}}$. This concludes the proof. 
\end{proof}

In maneuver-level cooperation, the evader actively participates in achieving the angle $\chi^{\star} \in \left[{\pi}/{2},{3\pi}/{2}\right]$ by sharing its engagement information with the defender, ensuring the interception of the pursuer from arbitrary initial engagement geometries. However, in certain applications, it may not be feasible for the evader to share all information due to various constraints, such as limited communication capabilities or safety concerns. Consequently, we now introduce an information-level cooperation scheme, where the evader only shares its maneuvering information with the defender. Based on this information, the defender's objective is to achieve the angle $\chi^\star$ within the specified time.

Having said that, in information-level cooperation, the evader independently performs maneuvers to evade the incoming pursuer, without direct coordination with the defender. In such scenarios, the evader may adopt a guidance strategy aimed at maximizing its relative separation from the pursuer. However, from the defender's perspective, it may be beneficial if the evader assists in capturing the pursuer, even though they are not actively coordinating. Toward this objective, the evader may employ a guidance strategy that renders it non-maneuvering with respect to the pursuer, effectively acting as a decoy. By presenting itself as a decoy, the evader entices the pursuer to continue on a non-maneuvering course under the belief that interception of the evader is imminent. Since the said approach reduces the pursuer’s evasive actions, this behavior is advantageous from the standpoint of the defender. It is also important to note that the pursuer becoming non-maneuvering is not a necessary condition for capturing the pursuer by the defender.

The evader can place itself as a decoy by nullifying its LOS rate relative to the pursuer, whose design is presented in the following theorem.
\begin{theorem}\label{thm:aE}
Consider the engagement geometry between the evader and the defender as shown in \Cref{fig:enggeo}. Let the evader's guidance law be
\begin{align}
    a_{\mathrm{E}} = \dfrac{r_{\mathrm{EP}}}{\cos \delta_{\mathrm{EP}}} \left(\dfrac{-2\dot{r}_{\mathrm{EP}}\dot{\lambda}_{\mathrm{EP}}}{r_{\mathrm{EP}}}  + \dot{h}_{\mathrm{E}}  + \mathcal{K}_1 \sign(\mathcal{S}_{\mathrm{E}}) \right) \label{eq:ae_passive}
\end{align}
for some $\mathcal{K}_1>\sup_{t\geq 0}\dfrac{a_{\mathrm{P}}^{\max}}{r_{\mathrm{EP}}}$, with
\begin{align}
\mathcal{S}_{\mathrm{E}}(t) = \dot{\lambda}_{\mathrm{EP}} + h_{\mathrm{E}}(t), \label{eq:se}
\end{align}
where $h_{\mathrm{E}}(t)$ is given by
\begin{equation}\label{eq:het}
    h_{\mathrm{E}}(t)=\begin{cases}
        \dfrac{h_{\mathrm{E}}(0)}{t_2^{k_3}}\left(t_2-t\right)^{k_3};& 0\leq t <t_2,\\
        \phantom{t-t}0; & t\geq t_2,
    \end{cases}
\end{equation}
for some {$k_3 \in \mathbb{N}$}, and $h_{\mathrm{E}}(0)$ is designed to ensure $\mathcal{S}_{\mathrm{E}}(0)=0$, given by $h_{\mathrm{E}}(0)=-\dot{\lambda}_{\mathrm{EP}}(0)$. Then, the LOS rate between the evader and the pursuer, $\dot{\lambda}_{\mathrm{EP}}$, converges to zero exactly at a prescribed time instant $t_2$, which can be assigned a priori, independent of the initial engagement geometry.
\end{theorem}
\begin{proof}
Consider a Lyapunov function candidate as, $\mathcal{V}_{2} = \frac{1}{2} \mathcal{S}_{\mathrm{E}}^2 $, whose time derivative can be obtained as
\begin{align}
\dot{\mathcal{V}}_{2} = \mathcal{S}_{\mathrm{E}} \dot{\mathcal{S}}_{\mathrm{E}} = \mathcal{S}_{\mathrm{E}} \left( \ddot{\lambda}_{\mathrm{EP}} + \dot{h}_{\mathrm{E}}  \right). \label{eq:v2_dot_1}
\end{align}
To obtain the value of $\ddot{\lambda}_{\mathrm{EP}}$, we differentiate \eqref{eq:lambdadotep} with respect to time and get
\begin{align}\label{eq:lambdaddotep1}
r_{\mathrm{EP}}\ddot{\lambda}_{\mathrm{EP}}+ \dot{r}_{\mathrm{EP}}\dot{\lambda}_{\mathrm{EP}}=v_{\mathrm{P}}
\cos \delta_{\mathrm{PE}} \dot{\delta}_{\mathrm{PE}}-v_{\mathrm{E}}
\cos \delta_{\mathrm{EP}} \dot{\delta}_{\mathrm{EP}}.
\end{align}
Using the relations $\dot{\delta}_{\mathrm{PE}}=\dot{\gamma}_{\mathrm{P}}-\dot{\lambda}_{\mathrm{EP}}$ and $\dot{\delta}_{\mathrm{EP}}=\dot{\gamma}_{\mathrm{E}}-\dot{\lambda}_{\mathrm{EP}}$, together with \eqref{eq:basic}, the expression in \eqref{eq:lambdaddotep1} becomes
\begin{align}
r_{\mathrm{EP}}\ddot{\lambda}_{\mathrm{EP}}=&~v_{\mathrm{P}}\cos \delta_{\mathrm{PE}} \left( \dfrac{a_{\mathrm{ P}}}{v_\mathrm{P}}-\dot{\lambda}_{\mathrm{EP}}\right) -v_{\mathrm{E}}\cos \delta_{\mathrm{EP}}  \left( \dfrac{a_{\mathrm{ E}}}{v_\mathrm{E}}-\dot{\lambda}_{\mathrm{EP}}\right) \nonumber\\
 &~-\dot{r}_{\mathrm{EP}}\dot{\lambda}_{\mathrm{EP}}.\label{eq:lambdaddotep2}
\end{align}
By collecting the similar terms together in \eqref{eq:lambdaddotep2}, we get
\begin{align*}
 \ddot{\lambda}_{\mathrm{EP}}=&~ \dfrac{\dot{\lambda}_{\mathrm{EP}}}{r_{\mathrm{EP}}}\left(-\dot{r}_{\mathrm{EP}}-v_{\mathrm{P}}\cos \delta_{\mathrm{PE}}+v_{\mathrm{E}}\cos \delta_{\mathrm{EP}}\right) + \dfrac{\cos \delta_{\mathrm{PE}}}{r_{\mathrm{EP}}}a_{\mathrm{P}} \nonumber\\
 &~- \dfrac{\cos \delta_{\mathrm{EP}}}{r_{\mathrm{EP}}}a_{\mathrm{ E}},
 \end{align*}
 which can be simplified using the relation in \eqref{eq:rdotep} to
 \begin{align}
  \ddot{\lambda}_{\mathrm{EP}}= &~ \dfrac{\dot{\lambda}_{\mathrm{EP}}}{r_{\mathrm{EP}}}\left(-\dot{r}_{\mathrm{EP}}- \dot{r}_{\mathrm{EP}}\right)+ \dfrac{\cos \delta_{\mathrm{PE}}}{r_{\mathrm{EP}}}a_{\mathrm{P}}- \dfrac{\cos \delta_{\mathrm{EP}}}{r_{\mathrm{EP}}}a_{\mathrm{ E}} \nonumber\\
  =&~ \dfrac{-2\dot{r}_{\mathrm{EP}}\dot{\lambda}_{\mathrm{EP}}}{r_{\mathrm{EP}}}+ \dfrac{\cos \delta_{\mathrm{PE}}}{r_{\mathrm{EP}}}a_{\mathrm{P}}- \dfrac{\cos \delta_{\mathrm{EP}}}{r_{\mathrm{EP}}}a_{\mathrm{ E}}. \label{eq:lambdaddotepfinal}
 \end{align}
By substituting the value of $\ddot{\lambda}_{\mathrm{EP}}$ from \eqref{eq:lambdaddotepfinal} into \eqref{eq:v2_dot_1}, one may obtain $\dot{\mathcal{V}}_{2}$ as 
\begin{align*}
\dot{\mathcal{V}}_{2} = \mathcal{S}_{\mathrm{E}} \left(\dfrac{-2\dot{r}_{\mathrm{EP}}\dot{\lambda}_{\mathrm{EP}}}{r_{\mathrm{EP}}}+ \dfrac{\cos \delta_{\mathrm{PE}}}{r_{\mathrm{EP}}}a_{\mathrm{P}}- \dfrac{\cos \delta_{\mathrm{EP}}}{r_{\mathrm{EP}}}a_{\mathrm{ E}} + \dot{h}_{\mathrm{E}}\right).
\end{align*}
which, on substituting the evader's guidance command from \eqref{eq:ae_passive}, becomes
\begin{align}
\dot{\mathcal{V}}_{2} = \mathcal{S}_{\mathrm{E}} &\left[\dfrac{-2\dot{r}_{\mathrm{EP}}\dot{\lambda}_{\mathrm{EP}}}{r_{\mathrm{EP}}}+ \dfrac{\cos \delta_{\mathrm{PE}}}{r_{\mathrm{EP}}}a_{\mathrm{P}} + \dot{h}_{\mathrm{E}}  - \dfrac{\cos \delta_{\mathrm{EP}} r_{\mathrm{EP}}}{r_{\mathrm{EP}}\cos \delta_{\mathrm{EP}}}  \right. \nonumber \\
&\left.\left(\dfrac{-2\dot{r}_{\mathrm{EP}}\dot{\lambda}_{\mathrm{EP}}}{r_{\mathrm{EP}}} - \dot{h}_{\mathrm{E}} -\mathcal{K}_1 \sign(\mathcal{S}_{\mathrm{E}}) \right) \right] . \label{eq:v2_dot_2}
\end{align}
After some algebraic simplifications, the expression in \eqref{eq:v2_dot_2} becomes
\begin{align}
\dot{\mathcal{V}}_{2} =&~ \mathcal{S}_{\mathrm{E}} \left(\dfrac{\cos \delta_{\mathrm{PE}}}{r_{\mathrm{EP}}}a_{\mathrm{P}}   -\mathcal{K}_1 \sign(\mathcal{S}_{\mathrm{E}})\right), \nonumber\\
\leq&~ - \left( \mathcal{K}_1 - \dfrac{1}{r_{\mathrm{EP}}} a_{\mathrm{P}}^{\max} \right) \lvert \mathcal{S}_{\mathrm{E}} \rvert \leq 0,
\end{align}
which implies that if $\mathcal{S}_{\mathrm{E}}(0)=0$, then under the evader's guidance command, the sliding mode will be maintained on the sliding manifold for all future times, that is, $\mathcal{S}_{\mathrm{E}}(t) \equiv 0$ $\forall\, t\geq 0$.
Since $\mathcal{S}_{\mathrm{E}}(t) \equiv 0$ $\forall\, t\geq 0$, implies $\dot{\lambda}_{\mathrm{EP}}  + h_{\mathrm{E}}=0$ from \eqref{eq:se}, which can be written using the relation $h_{\mathrm{E}}(0) = - \dot{\lambda}_{\mathrm{EP}}(0)$ as
\begin{align*}
\dot{\lambda}_{\mathrm{EP}} + \dfrac{h_{\mathrm{E}}(0)}{t_2^{k_3}}\left(t_2-t\right)^{k_3} = 0 \implies \dot{\lambda}_{\mathrm{EP}} = \dfrac{\dot{\lambda}_{\mathrm{EP}}(0) \left(t_2-t\right)^{k_3}}{t_2^{k_3}}.
\end{align*}
and thus at $t=t_2$, $\dot{\lambda}_{\mathrm{EP}} = 0$. This completes the proof.
\end{proof}

Once the evader lures the pursuer by becoming non-maneuvering with respect to the pursuer, our next objective is to design the defender's guidance strategy to capture the pursuer. Having said that, in the information-level cooperation, the defender aims to attain the angle $\chi^{\star}\in [\pi/2,3\pi/2]$, which in turn, will ensure the pursuer's capture by the virtue of \Cref{prop:interceptioncondition}. In the following theorem, we present the defender's guidance strategy.
 \begin{theorem}[Information-level Cooperation]\label{thm:aD}
     Consider the three-body engagement described using \eqref{eq:engementkinematics}, the dynamics of the error variable, \eqref{eq:ddotchifinal}, and the dual-layer sliding manifold in \eqref{eq:st} and \eqref{eq:sit}. If the defender-evader team uses \emph{information-level} cooperation, and the guidance strategy of the defender is designed as
     \begin{align}
         a_{\mathrm{D}} =&~ \left[ -\dfrac{2\dot{r}_{\mathrm{DP}}\dot{\lambda}_{\mathrm{DP}}}{r_{\mathrm{DP}}}+\dfrac{2\dot{r}_{\mathrm{DE}}\dot{\lambda}_{\mathrm{DE}}}{r_{\mathrm{ED}}}+\dfrac{\cos \delta_{\mathrm{ED}}}{r_{\mathrm{ED}}}a_{\mathrm{E}}-\dot{g}(t)+\dot{h}(t) \right.\nonumber\\
         &~\left. + \mathcal{K} \sign(\bar{\mathcal{S}}) \right]   \times \dfrac{r_{\mathrm{ED}}r_{\mathrm{DP}}}{\cos{\delta_{\mathrm{DE}}}r_{\mathrm{DP}} -\cos{\delta_{\mathrm{DP}}}r_{\mathrm{ED}}} ,\label{eq:aD_design}
     \end{align}
     for some $\mathcal{K}>\sup_{t\geq 0}\dfrac{a_{\mathrm{P}}^{\max}}{r_{\mathrm{DP}}}$, then the defender maneuvers such that it attains the desired angle $\chi^\star\in\left[{\pi}/{2},{3\pi}/{2}\right]$, within a time $t^\star$ prescribed prior to the engagement, regardless of the three-body initial engagement geometry. 
 \end{theorem}
\begin{proof}
We consider the same Lyapunov function candidate as in \Cref{thm:joint}, that is, $\mathcal{V}_{3} = \frac{1}{2}\bar{\mathcal{S}}^2$, whose derivative can be written using the expression in \eqref{eq:ddotchifinal} as
    \begin{align}
        \dot{\mathcal{V}}_{3} &=\bar{\mathcal{S}}\left[\dfrac{2\dot{r}_{\mathrm{ED}}\dot{\lambda}_{\mathrm{ED}}}{r_{\mathrm{ED}}} -\dfrac{2\dot{r}_{\mathrm{DP}}\dot{\lambda}_{\mathrm{DP}}}{r_{\mathrm{DP}}}
 + \dfrac{\cos\delta_{\mathrm{PD}}}{r_{\mathrm{DP}}}a_{\mathrm{P}}-\dot{g}+\dot{h}\right.\nonumber\\
 &~\left.+ \dfrac{\cos \delta_{\mathrm{ED}}}{r_{\mathrm{ED}}}a_\mathrm{E} - \left(\dfrac{\cos \delta_{\mathrm{DE}}}{r_{\mathrm{ED}}}+\dfrac{\cos \delta_{\mathrm{DP}}}{r_{\mathrm{DP}}}\right)a_\mathrm{D}\right]  \label{eq:v3_dot_1}
    \end{align}
    With the defender's guidance law chosen as in \eqref{eq:aD_design}, the expression in \eqref{eq:v3_dot_1} becomes
    \begin{align}
        \dot{\mathcal{V}}_{3} =&~\bar{\mathcal{S}}\left[ \dfrac{\cos\delta_{\mathrm{PD}}}{r_{\mathrm{DP}}}a_{\mathrm{P}} -\mathcal{K}\sign(\bar{\mathcal{S}}) \right]\nonumber\\
        \leq&-\left( \mathcal{K}-\dfrac{1}{r_{\mathrm{DP}}}a_{\mathrm{P}}^{\max} \right)\lvert \bar{\mathcal{S}}\rvert<0~\forall~ \bar{\mathcal{S}} \neq 0.
    \end{align}
    This implies that if $\bar{\mathcal{S}}(0)=0$, and the evader-defender team uses information-level cooperation with the defender's guidance strategy given in $\eqref{eq:aD_design}$, then the system's states will remain on the sliding manifold $\bar{\mathcal{S}}(t)=0~\forall~t \geq 0$. By following a similar procedure, one can show that the defender attains an angle $\chi^\star \in [\pi/2,3\pi/2]$ at an exact time instant $t^\star$.
\end{proof}
\begin{remark}
For simplicity of design, the same sliding manifolds and associated parameters as in the maneuver-level cooperation case are employed. However, one can choose these design parameters independently, depending upon the design specifications.
\end{remark}
\begin{remark}
In the {information-level} cooperation, the evader shares its maneuvering information with the defender. As a matter of fact, $a_{\mathrm{E}}$ in the \eqref{eq:aD_design} is readily available to the defender's guidance system. The defender's guidance system utilizes this information to compute the guidance command to intercept the pursuer.
\end{remark}

\section{Simulations}\label{sec:simulations}
We now demonstrate the performance of the proposed nonlinear guidance strategy using simulation for various three-body engagement scenarios. We assume $v_\mathrm{P}=v_\mathrm{D}=200$ m/s, whereas $v_\mathrm{E}=100$ m/s. In each case, we place the evader at the origin of the inertial frame at the beginning of the engagement. The initial positions of the agents in the trajectory plots are denoted using square markers, whereas an intercept is represented using an asterisk. In the following plots, the pursuer's trajectories and inputs are shown in red, while those of the evader and defender are shown in blue and green, respectively. In our simulations, the pursuer and the defender possess the same maneuverability capabilities, limited to $\pm 20$ g. On the other hand, the evader can only apply a maximum steering control of $10$ g in either direction. Here, g$=9.81$ m/s\textsuperscript{2} is the acceleration due to gravity. Thus, the pursuer and the defender are evenly matched in terms of both speed and acceleration capabilities. 
\subsection{Maneuver-level cooperation between evader and defender}
We first consider maneuver-level cooperation between evader and defender, that is, the evader-defender team jointly maneuvers to achieve the prescribed angle $\chi^\star\in [\pi/2,3\pi/2]$. The design parameters are chosen to be $t_1=3$ s, $t^\star=6$ s, $k_1=6$, $k_2=3$, $\Sigma_{\mathrm{E}}=\Sigma_{\mathrm{D}}=1$ and $\mathcal{K}=5$. Such choices of $t_1$ and $t^\star$ ensure that the sliding manifold converges to zero within 3 s, whereas the error vanishes within 6 s.

\begin{figure*}[!ht]
\centering
\begin{subfigure}{.25\linewidth}
    \centering
    \includegraphics[width=\linewidth]{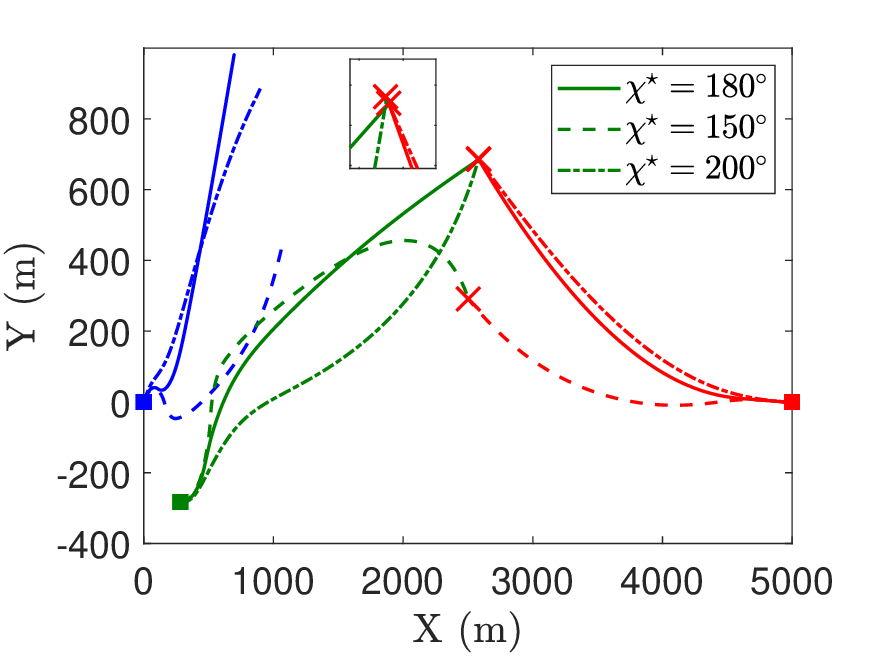}
    \caption{Trajectories.}
    \label{fig:moving_trajectory}
    \end{subfigure}%
    \begin{subfigure}{.25\linewidth}
    \centering
    \includegraphics[width=\linewidth]{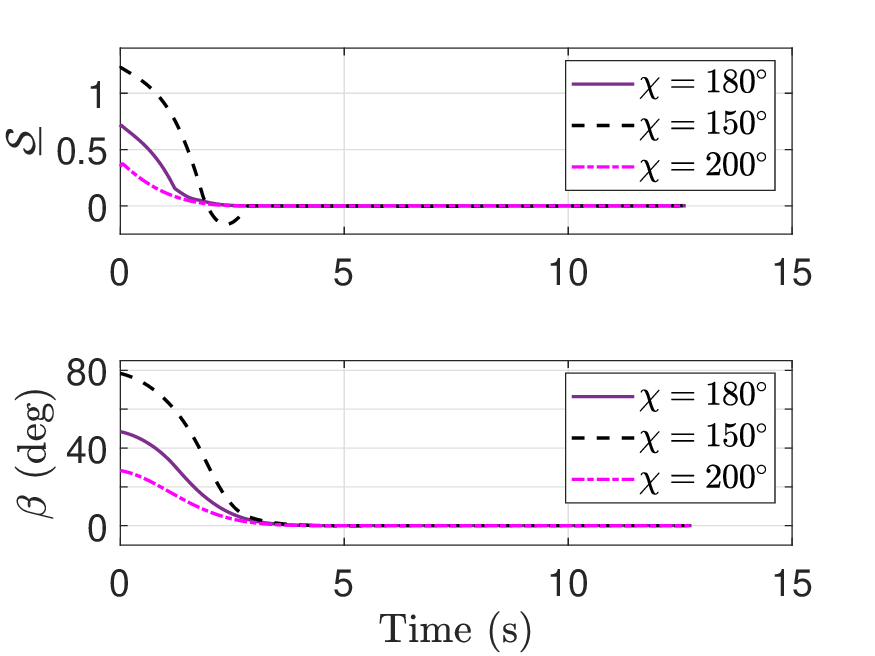}
    \caption{Sliding manifold and error.}
    \label{fig:moving_ss}
    \end{subfigure}%
    \begin{subfigure}{.25\linewidth}
    \centering
    \includegraphics[width=\linewidth]{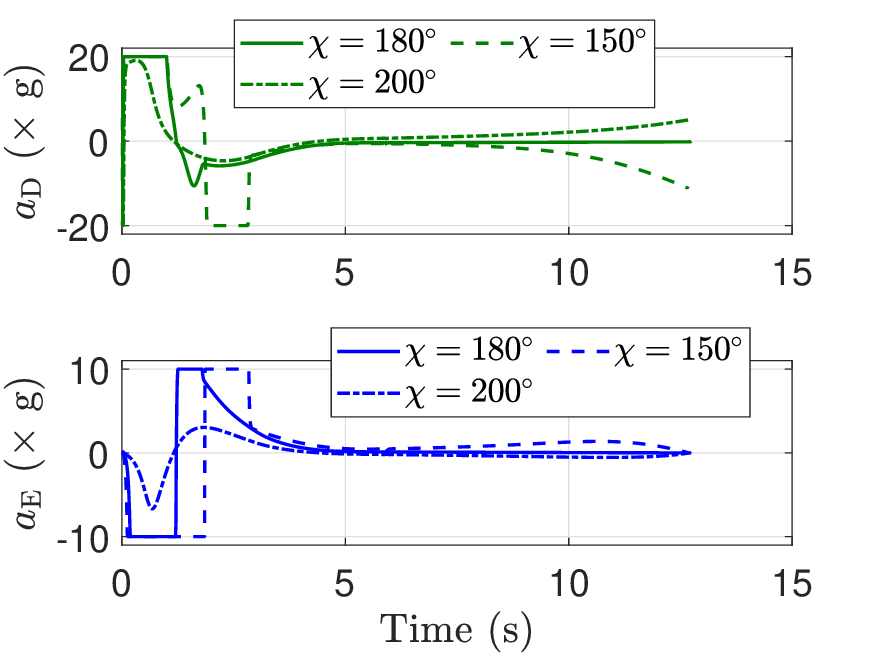}
    \caption{Steering controls $a_{\rm D}$ and $a_{\rm E}$.}
    \label{fig:moving_am}
    \end{subfigure}%
    \begin{subfigure}{.25\linewidth}
    \centering
    \includegraphics[width=\linewidth]{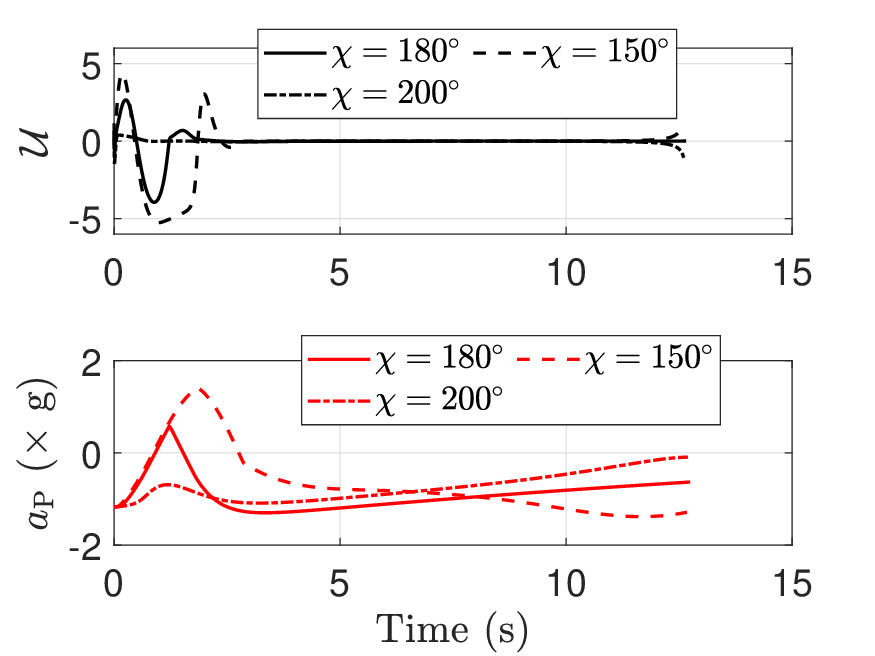}
    \caption{$\mathcal{U}$ and $a_{\rm P}$.}
    \label{fig:moving_u}
    \end{subfigure}
\caption{The defender intercepts the pursuer at various values of the angle $\chi^\star$.}
\label{fig:moving}
\end{figure*}

Consider a scenario where the pursuer is using a guidance strategy that is a function of the pursuer-evader LOS rate ($\dot{\lambda}_\mathrm{EP}$), for example, proportional-navigation guidance. For the pursuer, such a strategy may also be optimal in certain cases, as it aims to arrive on the collision course with minimum effort. The guidance command for the pursuer is given by $a_{\rm P} = - \mathscr{N} \dot{r}_{\mathrm{EP}} \dot{\lambda}_{\rm EP}$, where $\mathscr{N}=3$ is a constant. This situation is shown in  \Cref{fig:moving}. The defender and pursuer are initially located $400$ m and  $5000$ m radially apart from the evader, with respective LOS angles of $-45^\circ$ and $0^\circ$. The initial heading angles of the agents are $\gamma_{\mathrm{E}}=45^\circ$, $\gamma_{\mathrm{D}}=0^\circ$, and $\gamma_{\mathrm{P}}=180^\circ$. Under these settings, the initial value of the angle $\chi$ is $228^\circ$. It can be observed from \Cref{fig:moving_trajectory} that, based on different values of the desired $\chi^\star$, the agents maneuver differently. However, the pursuer is always intercepted before it can capture the evader. \Cref{fig:moving_ss} depicts the sliding manifold and the error profiles of various desired angles $\chi^\star$, which evidences that $\munderbar{\mathcal{S}}$ converges within 3 s and $\beta$ nullifies within 6 s, as expected. Therefore, regardless of the three-body initial engagement geometry, the evader-defender team maneuvers to ensure the pursuer's capture. \Cref{fig:moving_am,fig:moving_u} illustrate the various control efforts of the agents. It is important to note that the efforts of the evader and the defender (individually and jointly) have small magnitudes in the endgame.

\begin{figure*}[!ht]
\centering
	\begin{subfigure}{.25\linewidth}
    \centering
    \includegraphics[width=\linewidth]{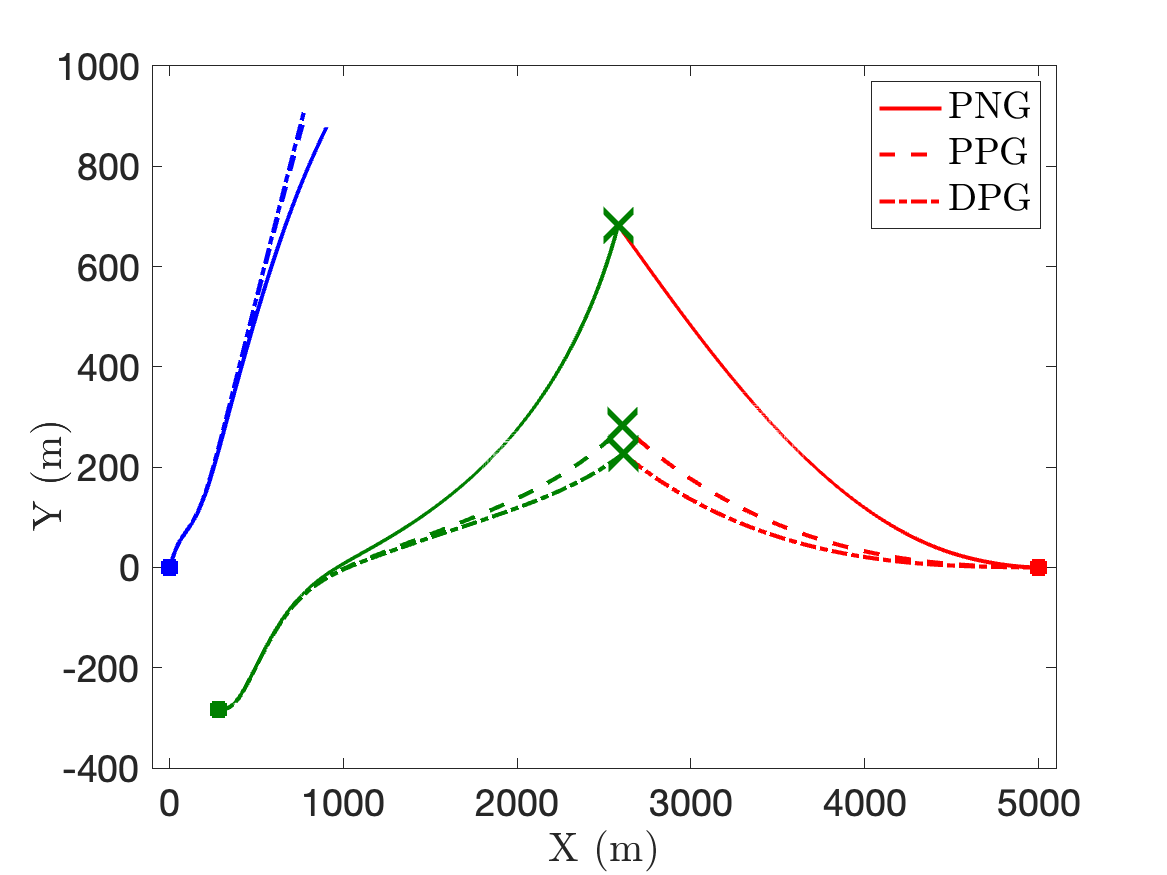}
    \caption{Trajectories.}
    \label{fig:moving_diffguidance_trajectory}
    \end{subfigure}%
    \begin{subfigure}{.25\linewidth}
    \centering
    \includegraphics[width=\linewidth]{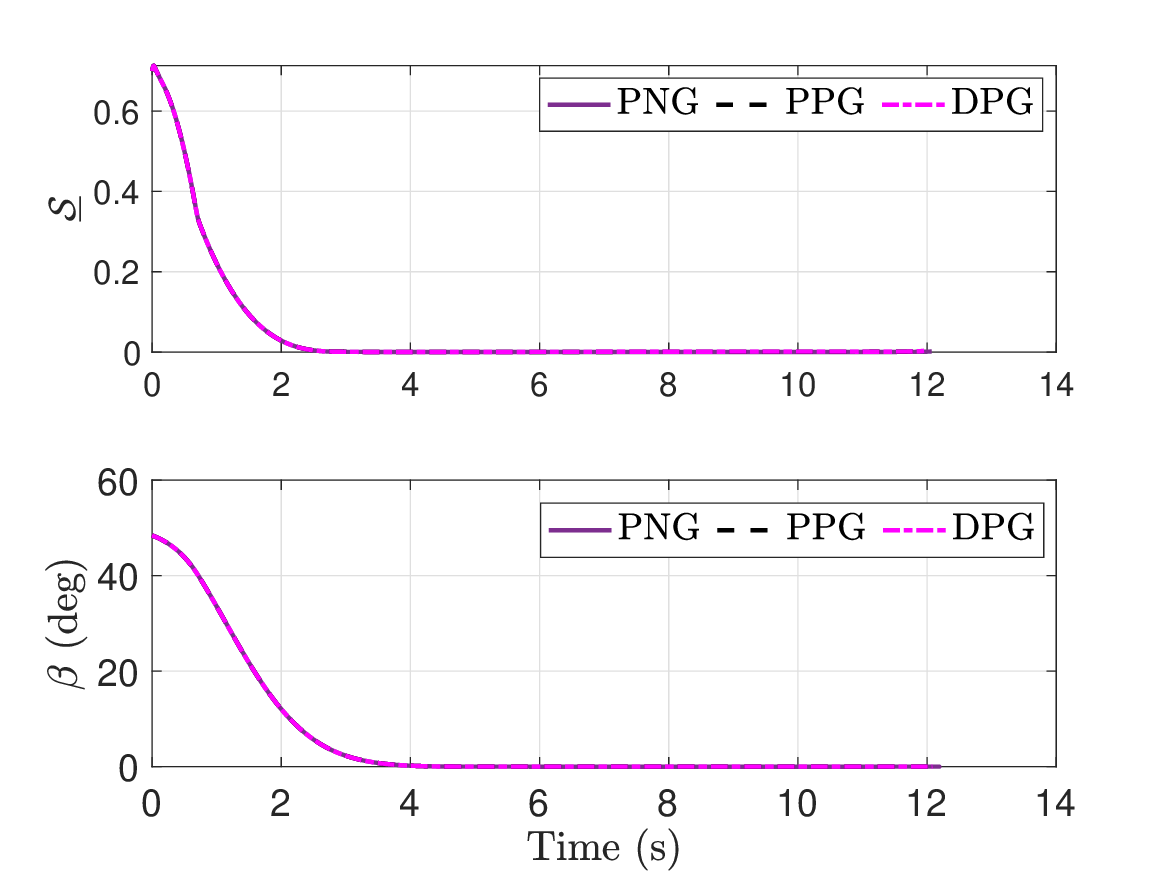}
    \caption{Sliding manifold and error.}
    \label{fig:moving_diffguidance_ss}
    \end{subfigure}%
    \begin{subfigure}{.25\linewidth}
    \centering
    \includegraphics[width=\linewidth]{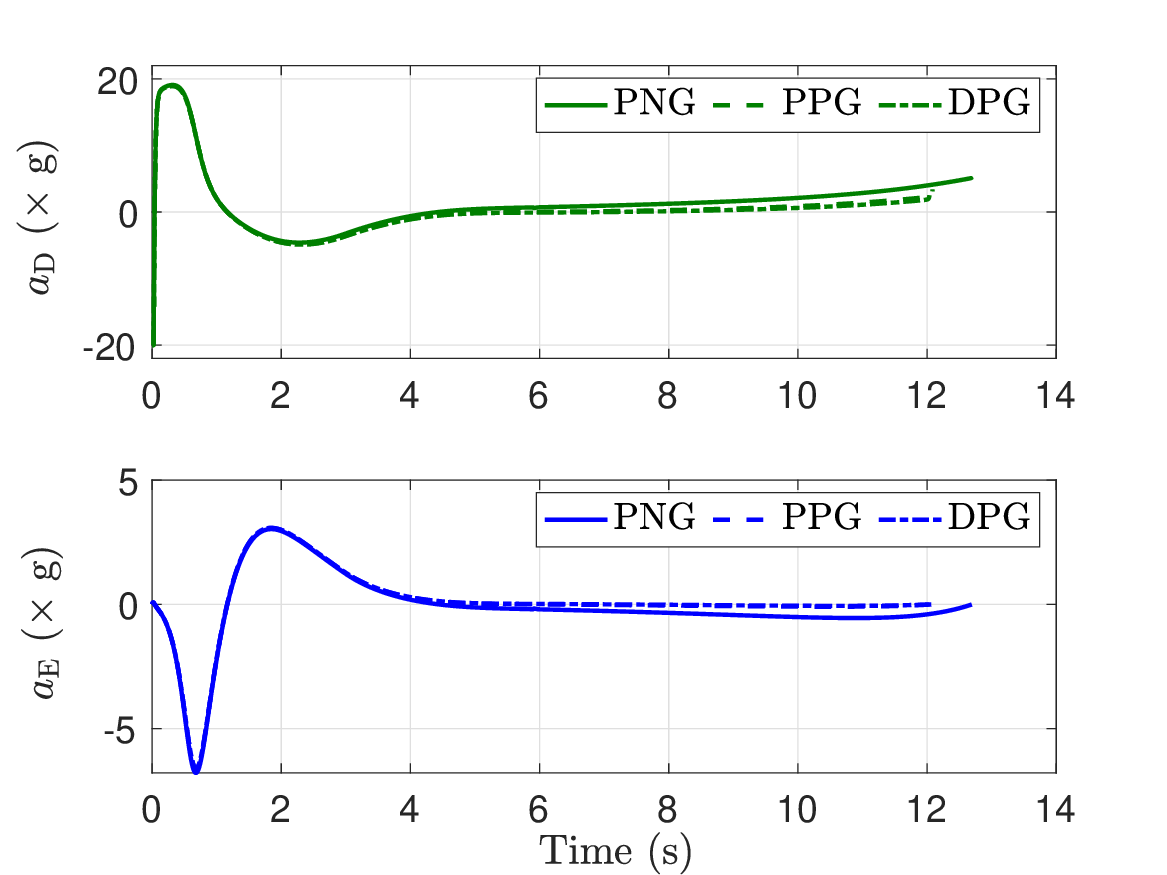}
    \caption{Steering controls $a_{\rm D}$ and $a_{\rm E}$.}
    \label{fig:moving_diffguidance_am}
    \end{subfigure}%
    \begin{subfigure}{.25\linewidth}
    \centering
    \includegraphics[width=\linewidth]{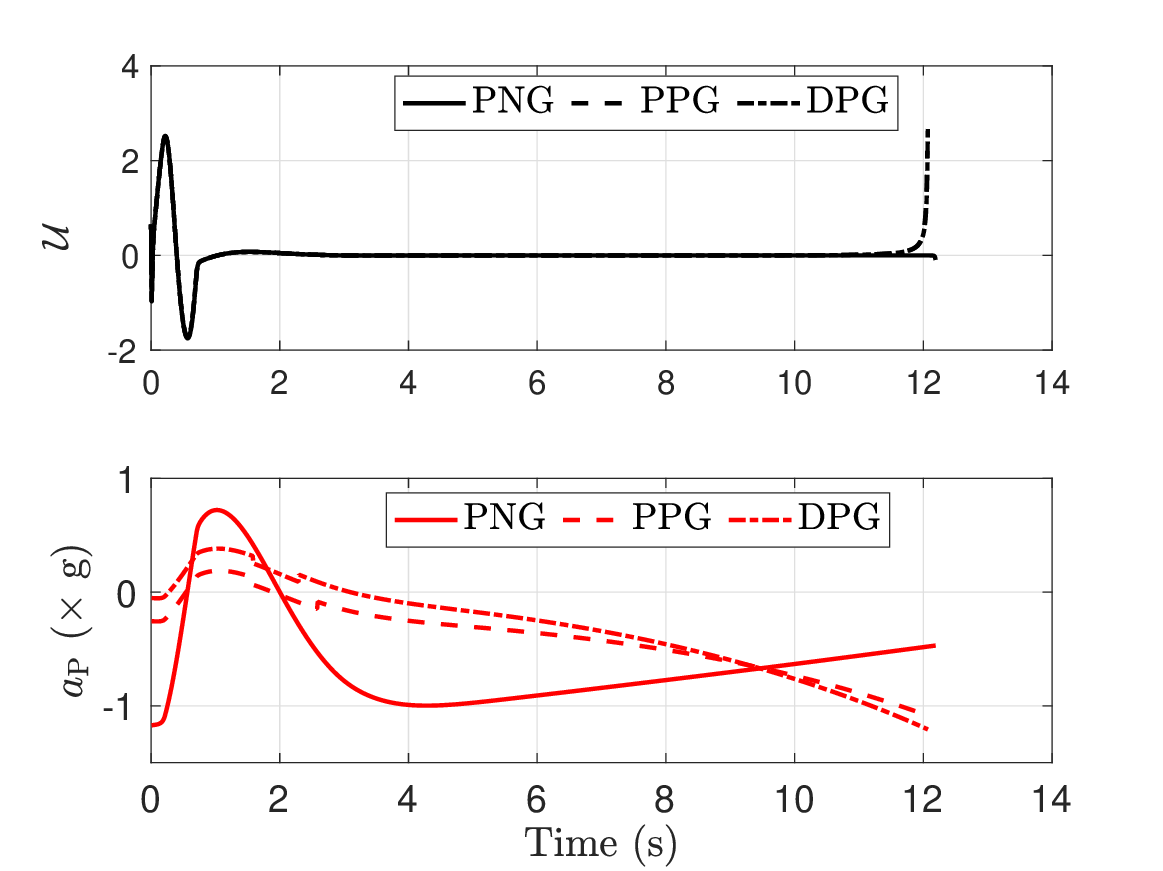}
        \caption{$\mathcal{U}$ and $a_{\rm P}$.}
    \label{fig:moving_diffguidance_u}
    \end{subfigure}
\caption{The pursuer uses different guidance strategies.}
\label{fig:moving_diffguidance}
\end{figure*}

Since the proposed defense strategy is independent of the pursuer's maneuver, the defender is able to capture the pursuer even if it uses a different guidance law. \Cref{fig:moving_diffguidance} portrays various cases when the pursuer uses proportional-navigation guidance (PNG), pure pursuit guidance (PPG), and deviated pursuit guidance (DPG) to capture the evader. The guidance command for PNG, PPG, and DPG are chosen to be $a_{\rm P} = - \mathscr{N} \dot{r}_{\mathrm{EP}} \dot{\lambda}_{\rm EP}$, $a_{\rm P} =  v_{\mathrm{P}} \dot{\lambda}_{\rm EP}- 0.1\left(\gamma_{\mathrm{P}} -\lambda_{\mathrm{EP}} \right)$, and $a_{\rm P} =  v_{\mathrm{P}} \dot{\lambda}_{\rm EP}- 0.1\left(\gamma_{\mathrm{P}} -\lambda_{\mathrm{EP}} - \delta \right)$, respectively with $\mathscr{N}=3$ and $\delta=20^{\circ}$. For a fixed angle, $\chi^\star=180^\circ$, the defender always intercepts the pursuer midway and safeguards the evader. In \Cref{fig:moving_diffguidance}, the initial conditions are kept the same as those in the previous case, except $\chi^\star=180^\circ$. We also observe similar behaviors of the sliding manifold, error, and the agents' control efforts.   

\begin{figure*}[!ht]
\centering
	\begin{subfigure}{.327\linewidth}
    \centering
    \includegraphics[width=\linewidth]{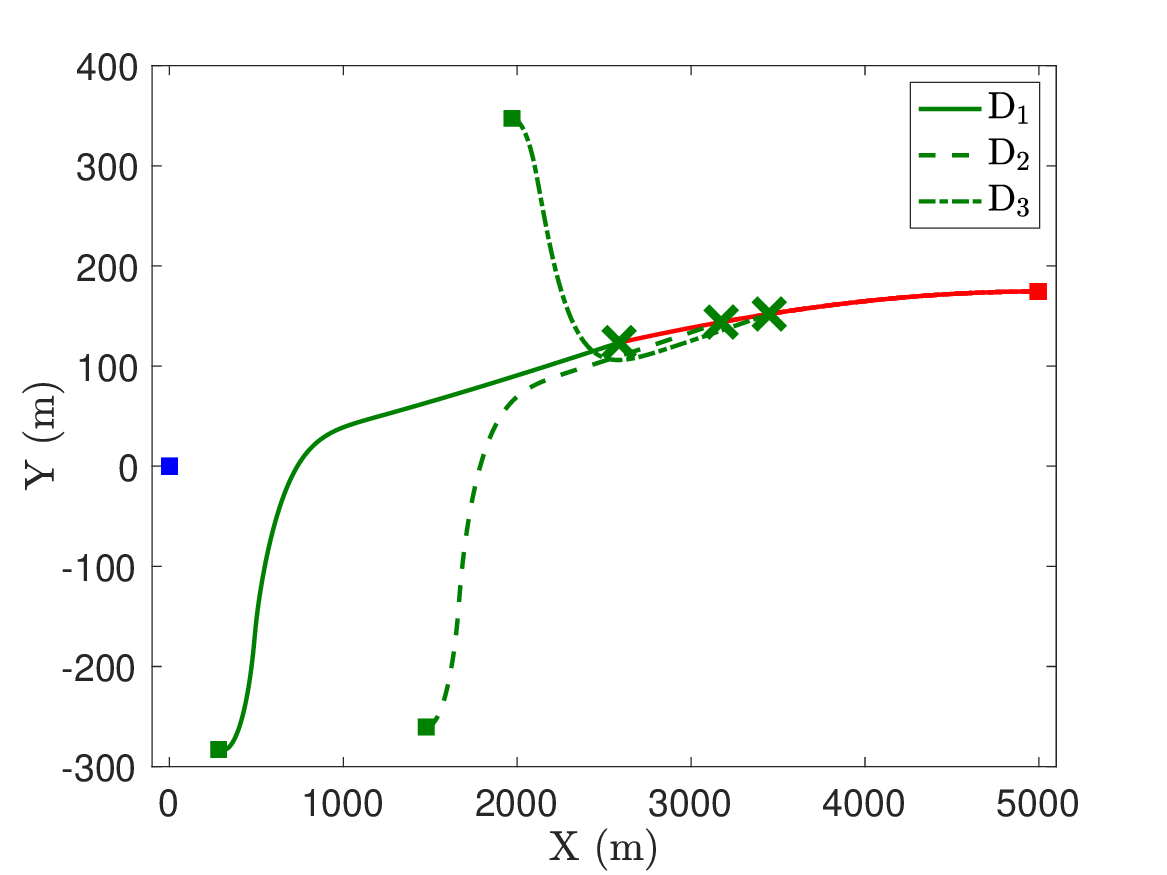}
    \caption{Trajectories.}
    \label{fig:stationary_trajectory}
    \end{subfigure}
     \begin{subfigure}{.327\linewidth}
    \centering
    \includegraphics[width=\textwidth]{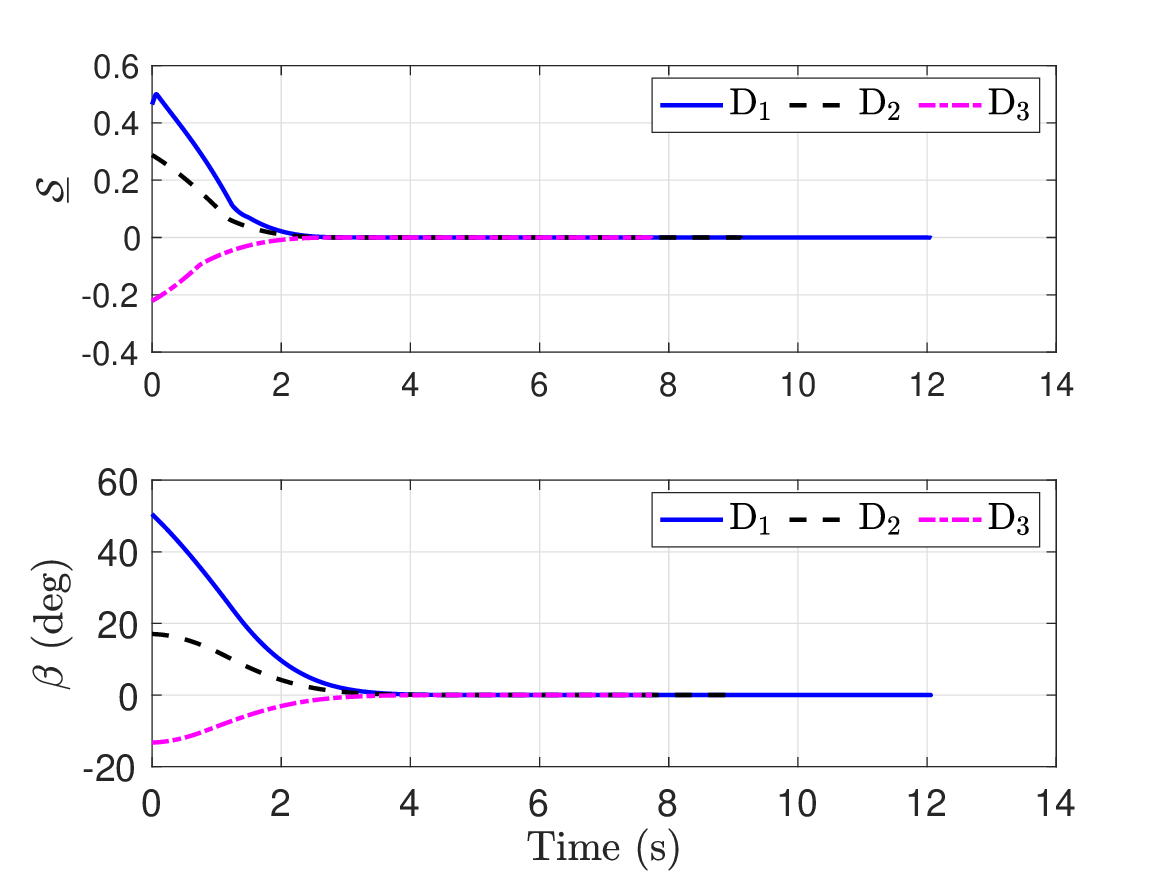}
    \caption{Sliding manifold and  error.}
    \label{fig:stationary_ss}
    \end{subfigure}
    \begin{subfigure}{.327\linewidth}
    \centering
    \includegraphics[width=\linewidth]{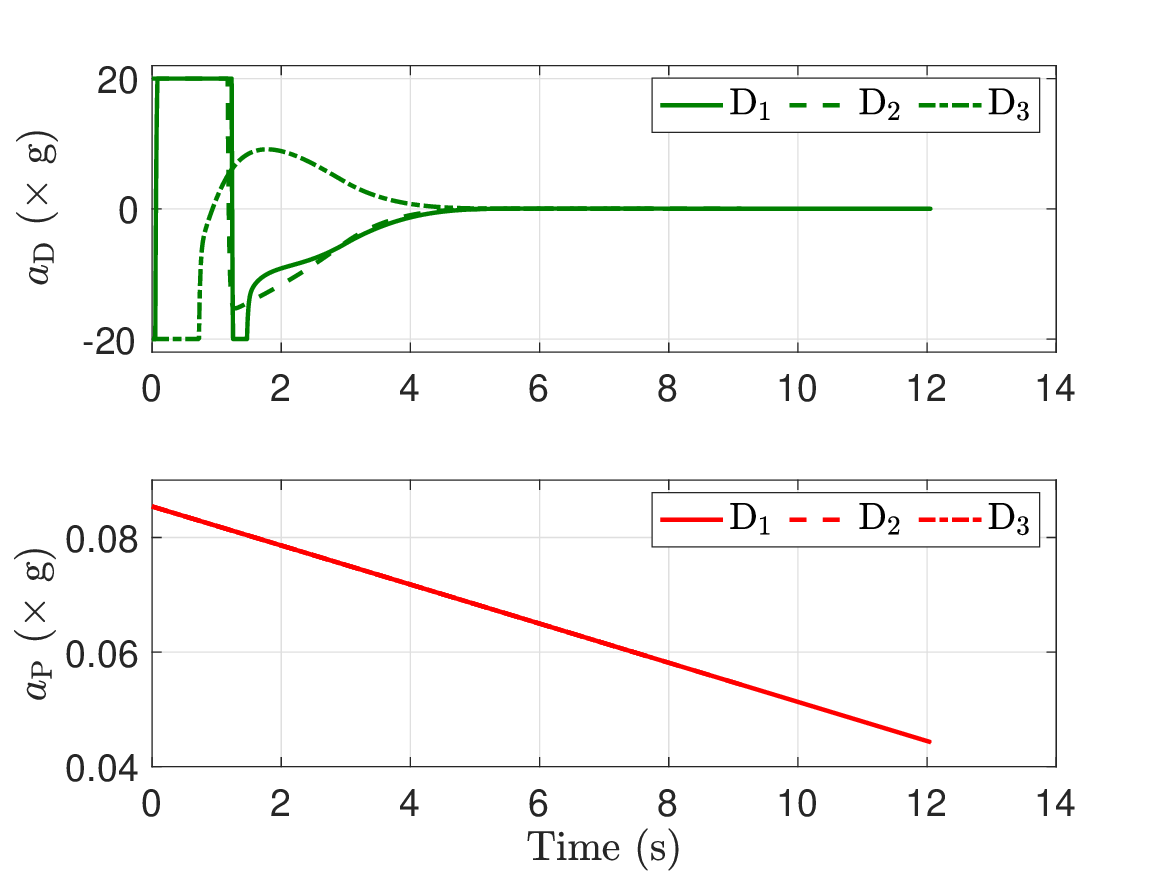}
    \caption{Steering controls $a_{\rm D}$ and $a_{\rm P}$.}
    \label{fig:stationary_am}
    \end{subfigure}
\caption{The defender safeguards the stationary evader from different initial configurations.}
\label{fig:stationary_evader}
\end{figure*}

\begin{figure*}[!ht]
\centering
	\begin{subfigure}{.327\linewidth}
    \centering
    \includegraphics[width=\linewidth]{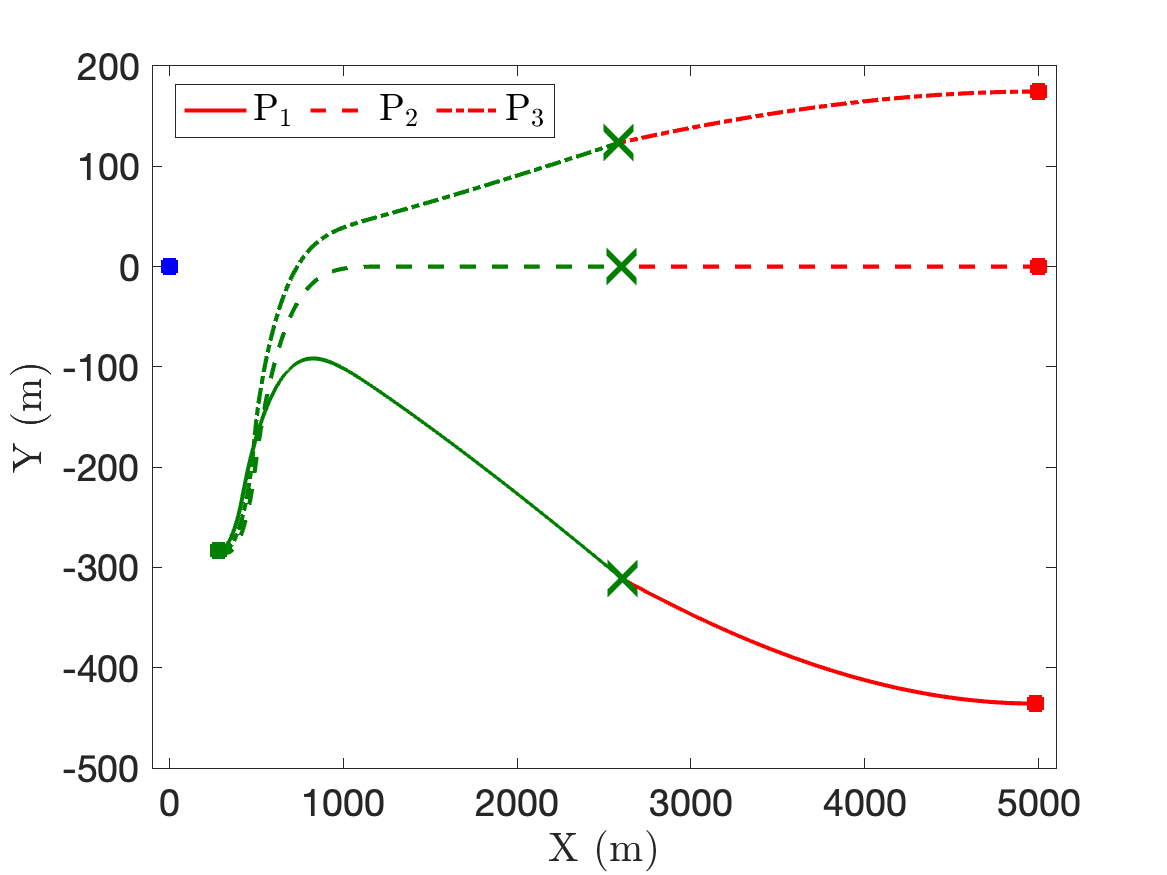}
    \caption{Trajectories.}
    \label{fig:stationary_diffattacker_trajectory}
    \end{subfigure}
        \begin{subfigure}{.327\linewidth}
    \centering
    \includegraphics[width=\linewidth]{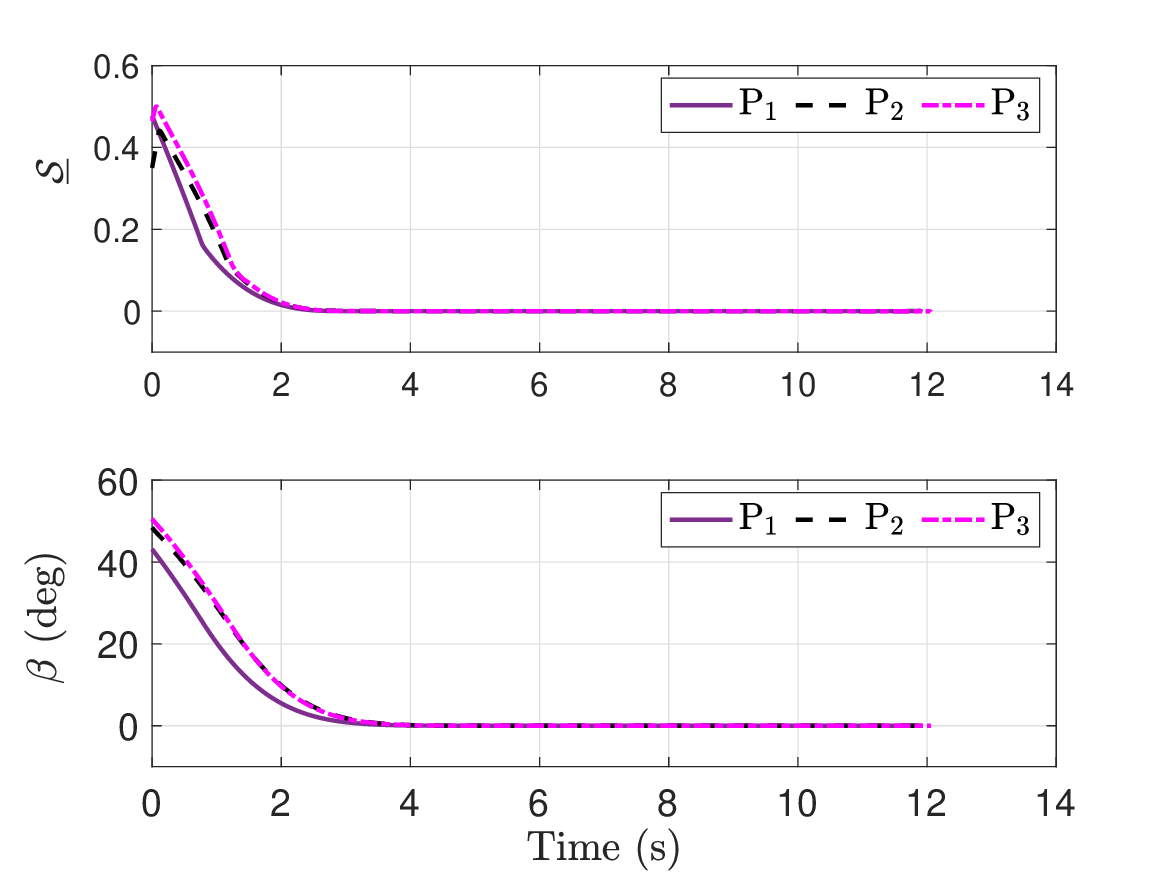}
    \caption{Sliding manifold and  error.}
    \label{fig:stationary_diffattacker_ss}
    \end{subfigure}
    \begin{subfigure}{.327\linewidth}
    \centering
    \includegraphics[width=\linewidth]{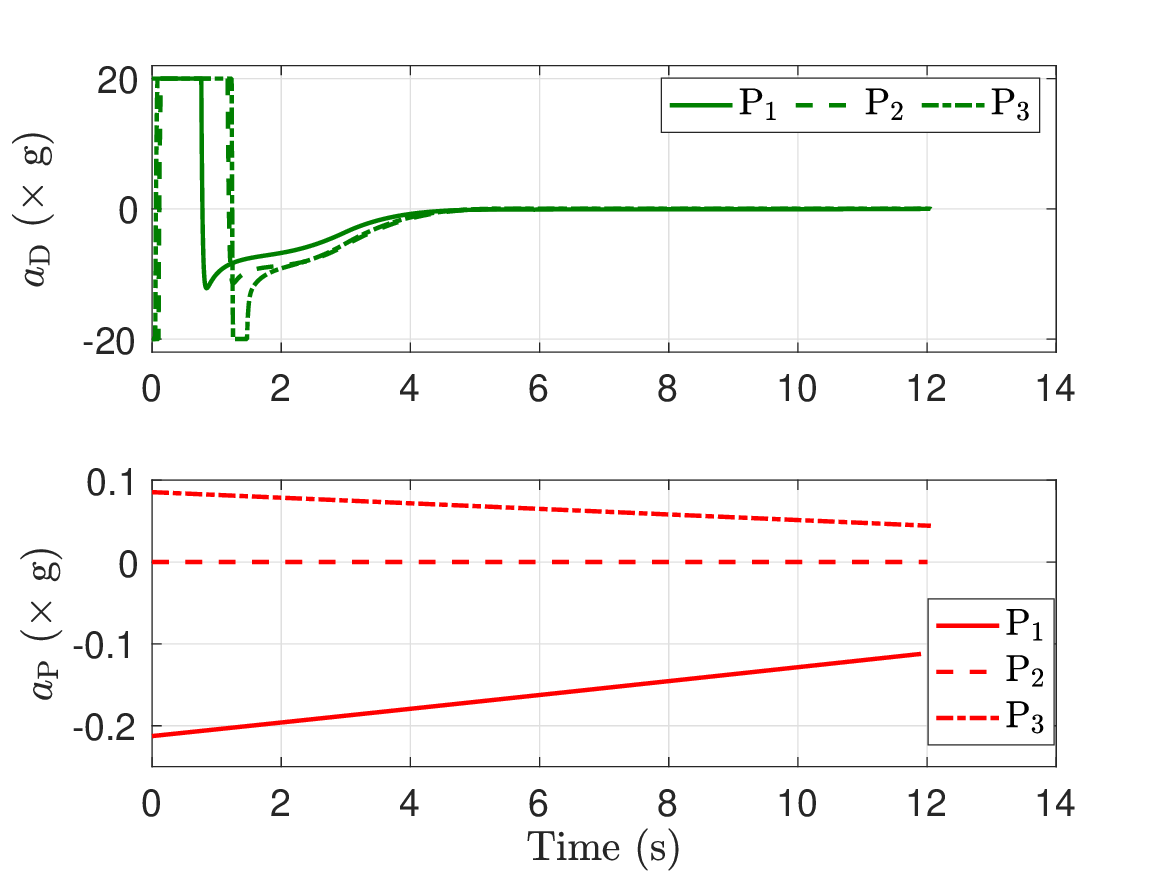}
    \caption{Steering controls of defender and pursuer.}
    \label{fig:stationary_diffattacker_am}
    \end{subfigure}
\caption{The defender safeguards the stationary evader from the same initial configuration.}
\label{fig:stationary_diffattacker}
\end{figure*}

We now consider cases when the evader is stationary, so the pursuer can head directly toward it with an optimal effort once any heading angle errors vanish. In \Cref{fig:stationary_evader}, the pursuer starts at a distance of 5000 m from the evader with a LOS angle of $2^\circ$. The defender, on the other hand, starts at three different positions in three different cases. With respect to the evader, the defender is at a relative separation of $400$ m, $1500$ m, and $2000$ m with LOS angles of $-45^\circ$, $-10^\circ$, and $10^\circ$, respectively. The defender aims to attain $\chi^\star=180^\circ$ in each case to arrive directly between the evader and the pursuer such that the latter's capture can be guaranteed. We notice that even if the evader does not maneuver, the defender maneuvers accordingly to satisfy the specific geometrical conditions for a constant $\chi^\star$, thereby guaranteeing pursuit-evasion. On the other hand, the situation in \Cref{fig:stationary_diffattacker} assumes a fixed position of the defender ($r_\mathrm{DE}=400$ m, $\lambda_\mathrm{DE}=-45^\circ$), whereas the pursuer is located at different positions in different cases. The initial configurations of the pursuer relative to the evader are $r_\mathrm{EP}=5000$ m with LOS angles of $-5^\circ, 0^\circ,$ and $2^\circ$, respectively. Once again, we observe that the fixed angle $\chi^\star$ is achieved by the defender in $t^\star=6$ s and maintained thereafter to ensure the pursuer's capture. 

\begin{figure*}[ht!]
\centering
\begin{subfigure}{.25\linewidth}
    \centering
    \includegraphics[width=\linewidth]{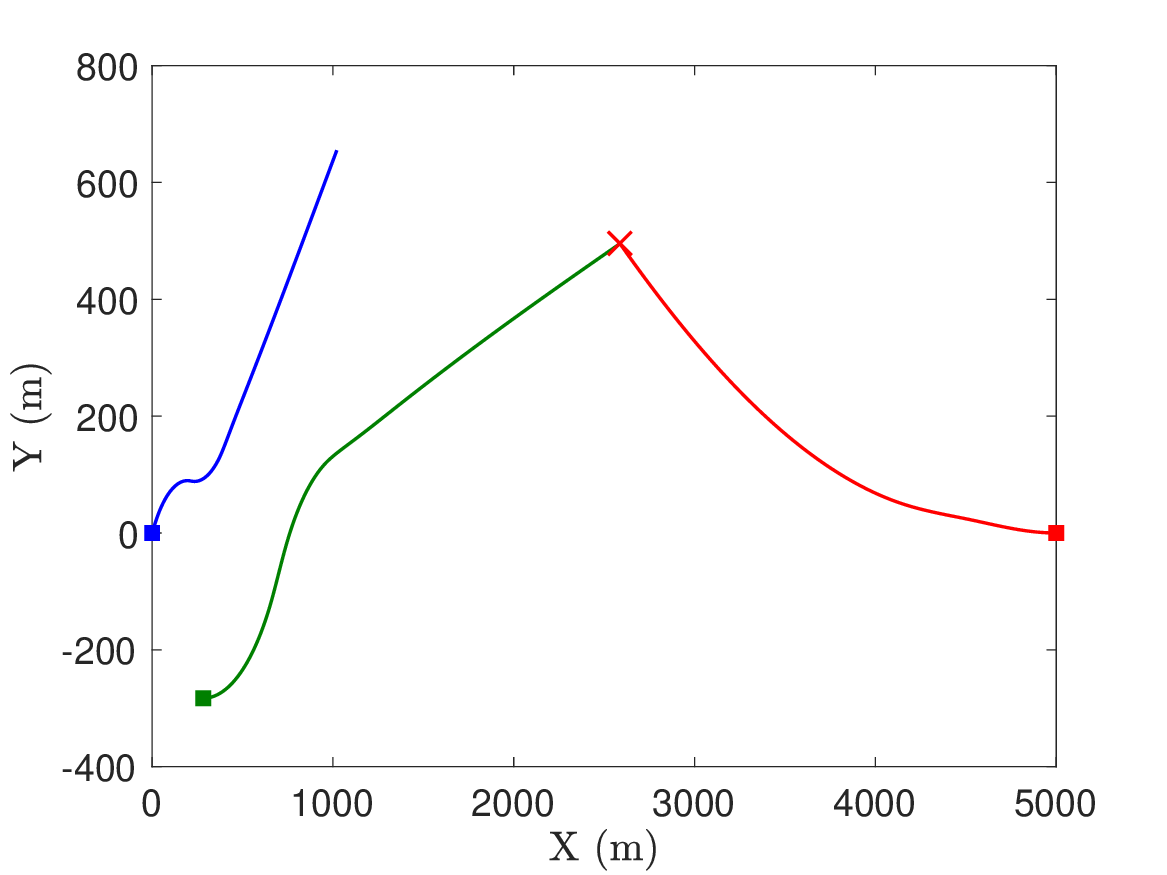}
    \caption{Trajectories.}
    \label{fig:trajectory_reduced_amax}
    \end{subfigure}%
    \begin{subfigure}{.25\linewidth}
    \centering
    \includegraphics[width=\linewidth]{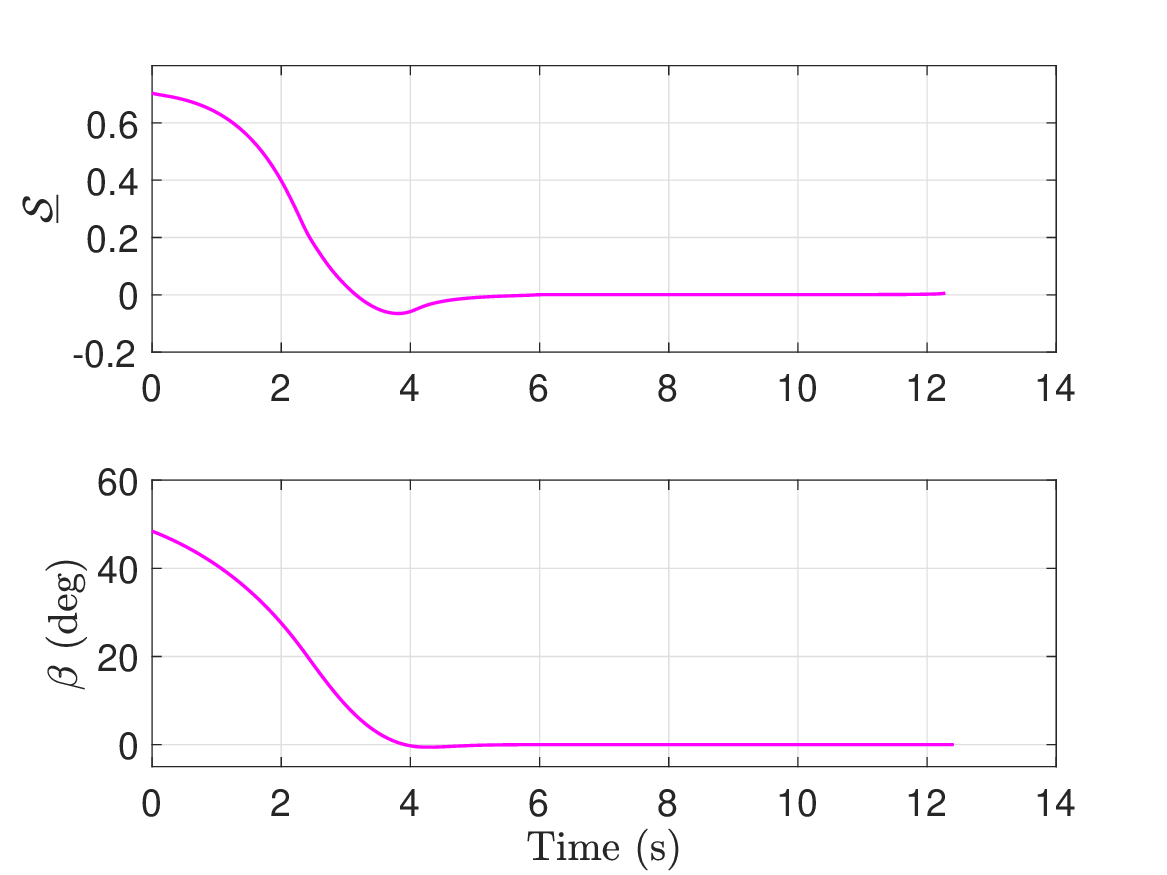}
    \caption{Sliding manifold and  error.}
    \label{fig:s_beta_reduced_amax}
    \end{subfigure}%
    \begin{subfigure}{.25\linewidth}
    \centering
    \includegraphics[width=\linewidth]{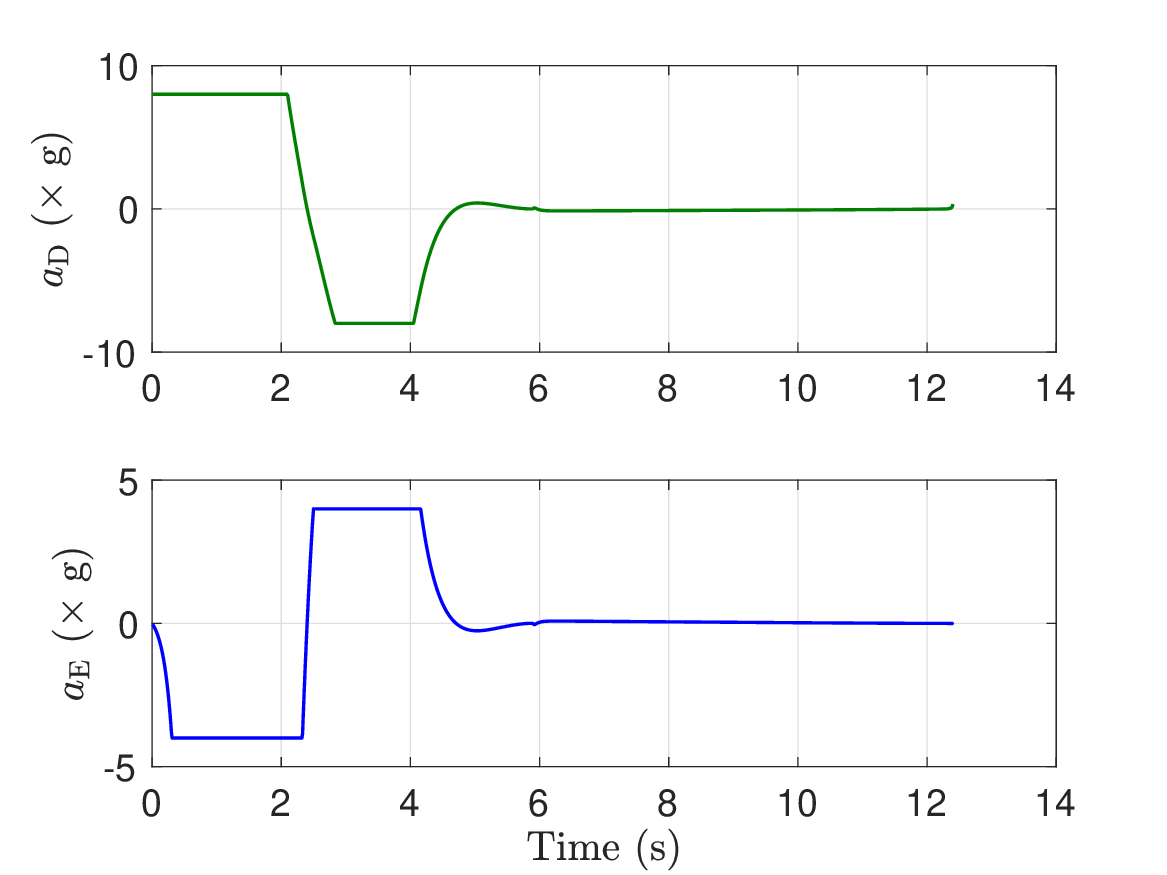}
    \caption{Steering controls $a_{\rm D}$ and $a_{\rm E}$.}
    \label{fig:ad_ae_reduced_amax}
    \end{subfigure}%
    \begin{subfigure}{.25\linewidth}
    \centering
    \includegraphics[width=\linewidth]{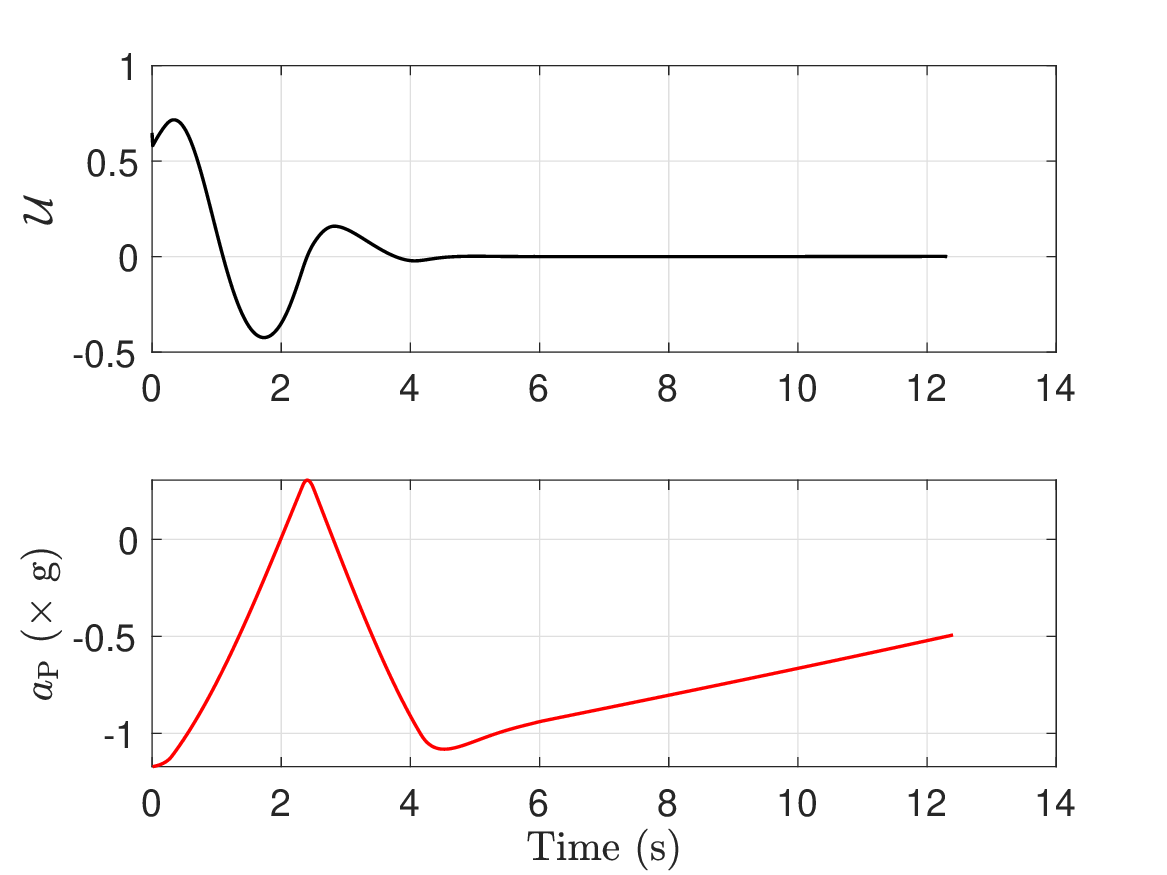}
    \caption{$\mathcal{U}$ and $a_{\rm P}$.}
    \label{fig:u_ap_reduced_amax}
    \end{subfigure}
\caption{The defender intercepts the pursuer with reduced maximum acceleration}.
\label{fig:reduced_amax}
\end{figure*}

We now demonstrate the performance of the proposed strategy when the maximum steering controls of the defender and pursuer are chosen to be $8$ g, whereas the maximum steering control input for the evader is chosen to be $4$ g. While the other engagement settings are kept the same as for $\chi^{\star}=180^{\circ}$, we depict the performance under reduced maximum capability through \Cref{fig:reduced_amax}. It can be observed that the defender intercepts the pursuer by following a similar behavior as when maximum accelerations are higher. This further bolsters the arguments presented in this work that pursuit-evasion is guaranteed regardless of the three-body initial engagement geometry and the knowledge of the pursuer's strategy.
\begin{figure*}[!ht]
\centering
\begin{subfigure}{.25\linewidth}
    \centering
    \includegraphics[width=\linewidth]{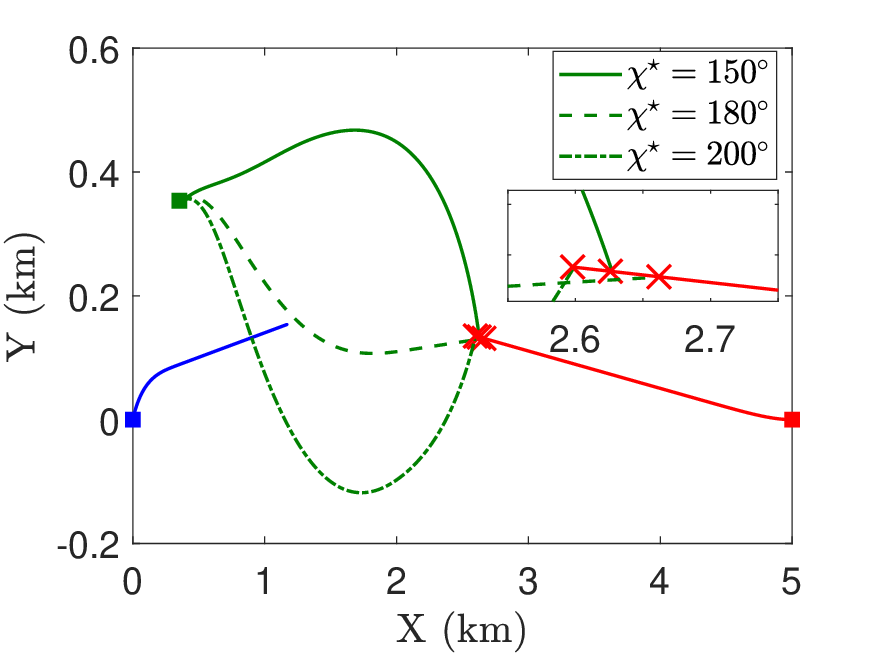}
    \caption{Trajectories.}
    \label{fig:p_diff_chi_trajectory}
    \end{subfigure}%
    \begin{subfigure}{.25\linewidth}
    \centering
    \includegraphics[width=\linewidth]{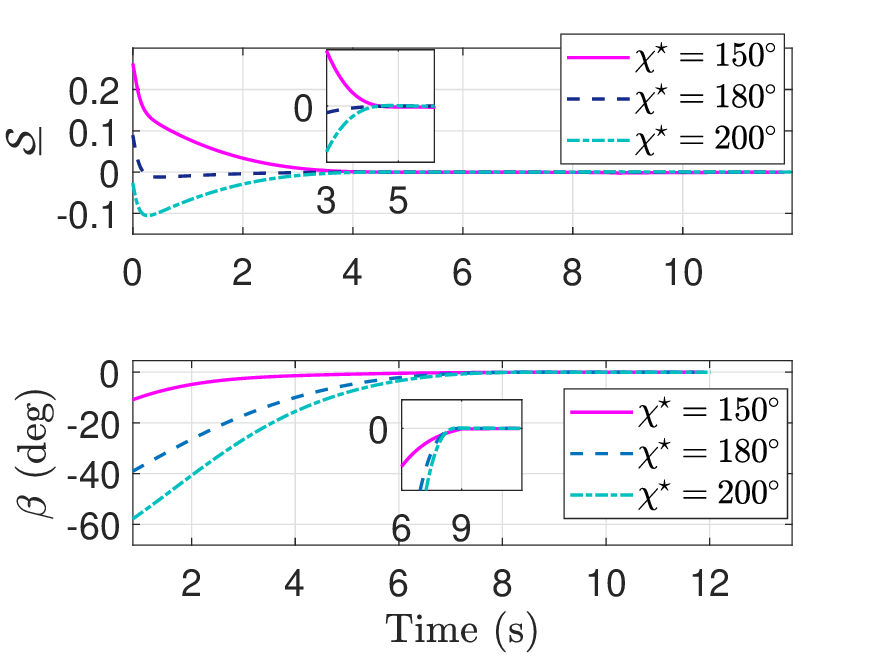}
    \caption{Sliding manifold and error.}
    \label{fig:p_diff_chi_s_beta}
    \end{subfigure}%
    \begin{subfigure}{.25\linewidth}
    \centering
    \includegraphics[width=\linewidth]{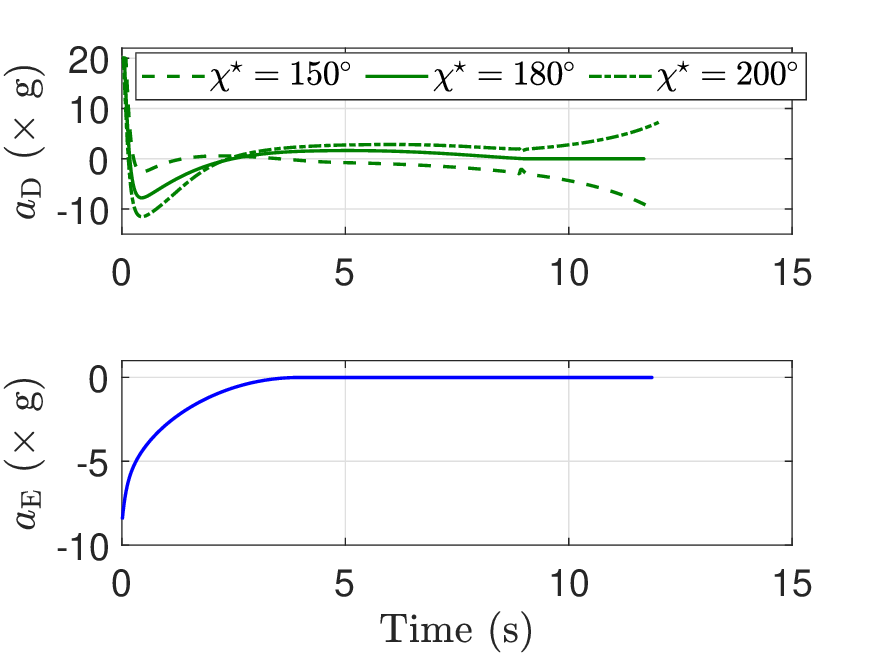}
    \caption{Steering controls $a_{\mathrm{D}}$ and $a_{\mathrm{E}}$.}
    \label{fig:p_diff_chi_ad_ae}
    \end{subfigure}%
    \begin{subfigure}{.25\linewidth}
    \centering
    \includegraphics[width=\linewidth]{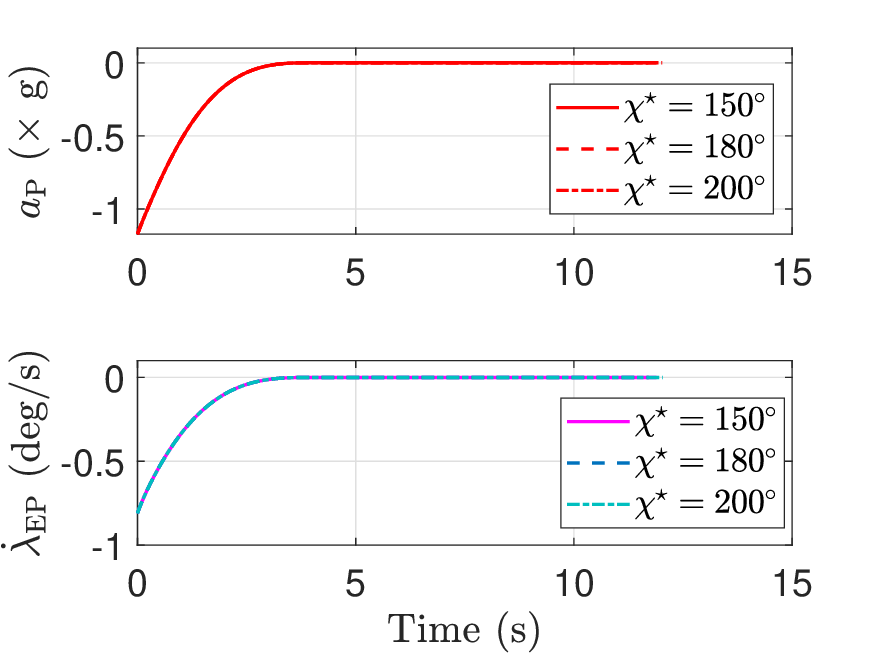}
    \caption{$a_{\mathrm{P}}$ and $\dot{\lambda}_{\mathrm{EP}}$.}
    \label{fig:p_diff_chi_ap_dotlambda}
    \end{subfigure}
\caption{The defender intercepts the pursuer at various values of the angle $\chi^\star$.}
\label{fig:p_diff_chi}
\end{figure*}
\begin{figure*}[!ht]
\centering
\begin{subfigure}{.25\linewidth}
    \centering
    \includegraphics[width=\linewidth]{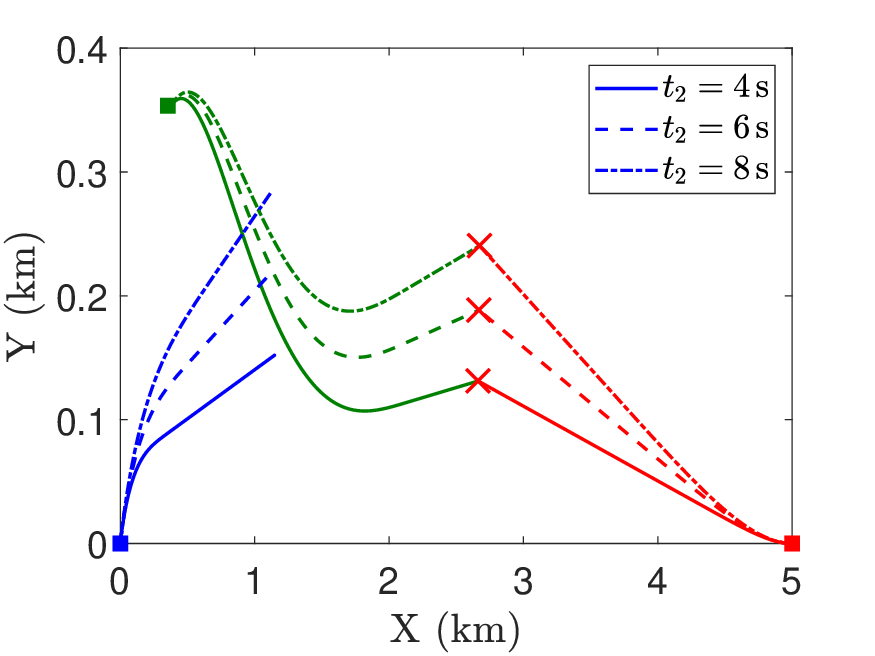}
    \caption{Trajectories.}
    \label{fig:p_diff_t2_trajectory}
    \end{subfigure}%
    \begin{subfigure}{.25\linewidth}
    \centering
    \includegraphics[width=\linewidth]{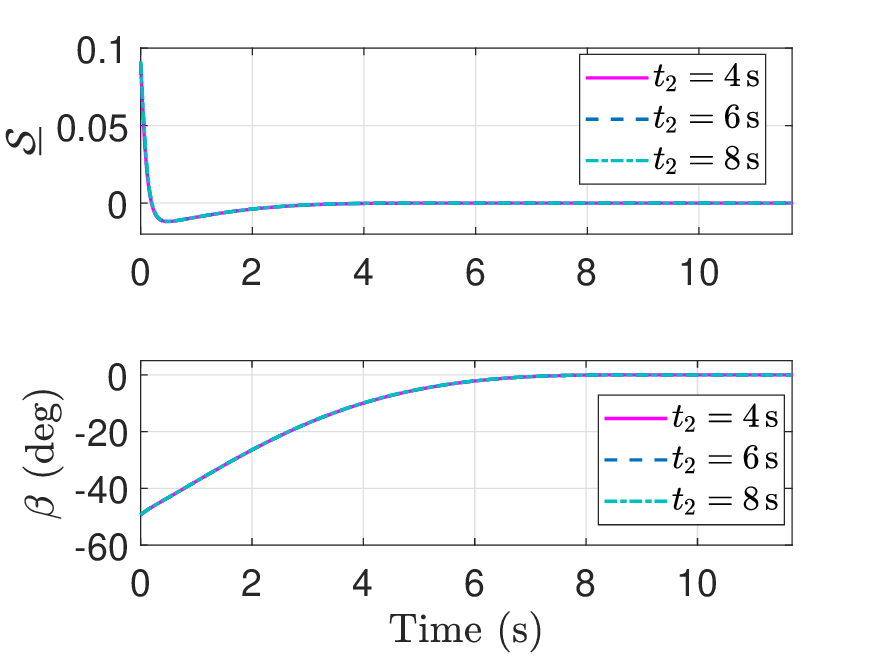}
    \caption{Sliding manifold and error.}
    \label{fig:p_diff_t2_s_beta}
    \end{subfigure}%
    \begin{subfigure}{.25\linewidth}
    \centering
    \includegraphics[width=\linewidth]{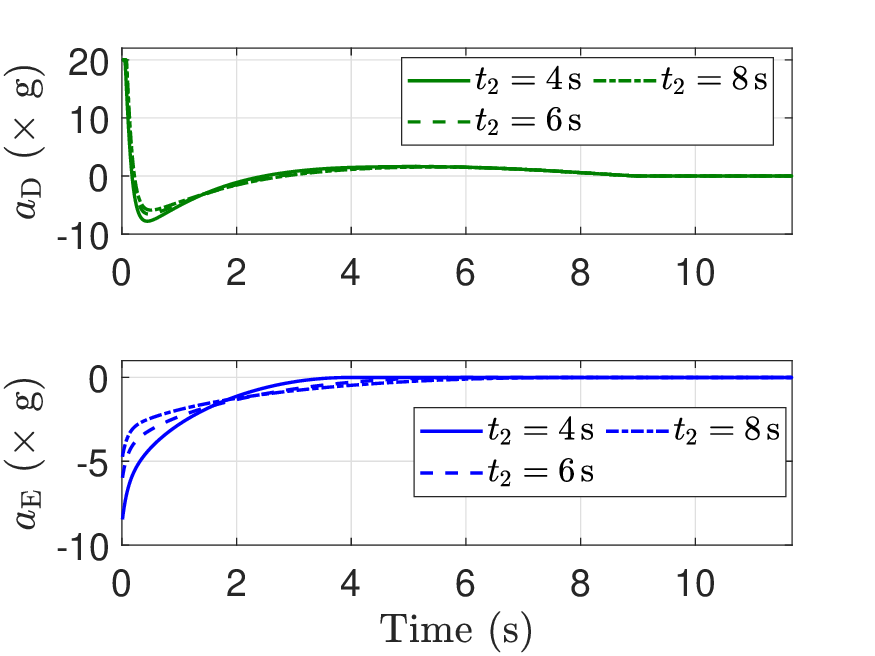}
    \caption{Steering controls $a_{\mathrm{D}}$ and $a_{\mathrm{E}}$.}
    \label{fig:p_diff_t2_ad_ae}
    \end{subfigure}%
    \begin{subfigure}{.25\linewidth}
    \centering
    \includegraphics[width=\linewidth]{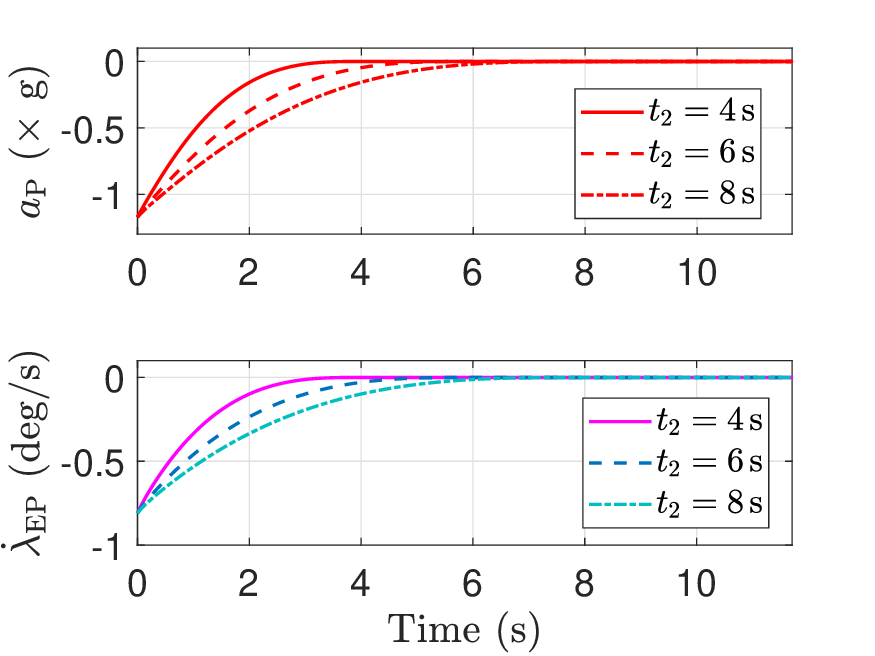}
    \caption{$a_{\mathrm{P}}$ and $\dot{\lambda}_{\mathrm{EP}}$.}
    \label{fig:p_diff_t2_ap_dotlambda}
    \end{subfigure}
\caption{The defender intercepts the pursuer for various values of time $t_2$.}
\label{fig:p_diff_t2}
\end{figure*}
\begin{figure*}[!ht]
\centering
\begin{subfigure}{.25\linewidth}
    \centering
    \includegraphics[width=\linewidth]{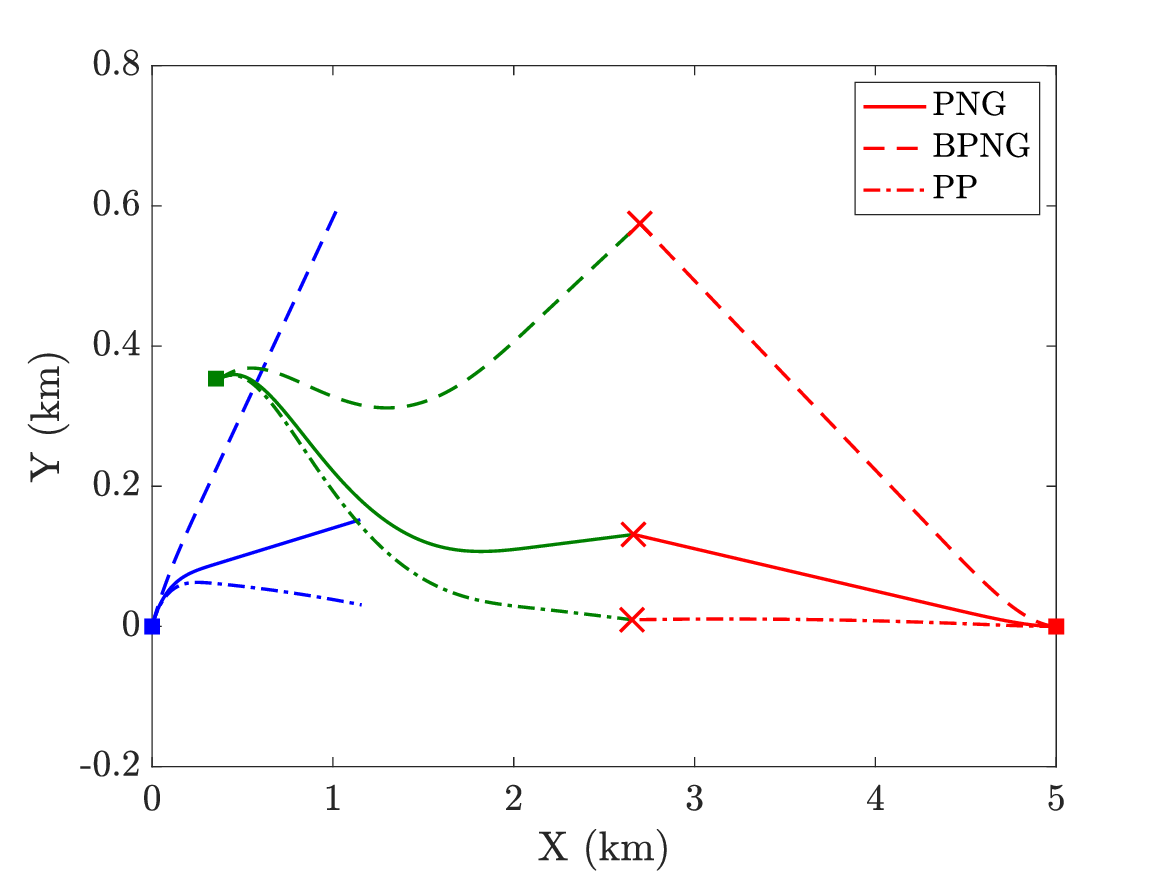}
    \caption{Trajectories.}
    \label{fig:p_diff_guid_path}
    \end{subfigure}%
    \begin{subfigure}{.25\linewidth}
    \centering
    \includegraphics[width=\linewidth]{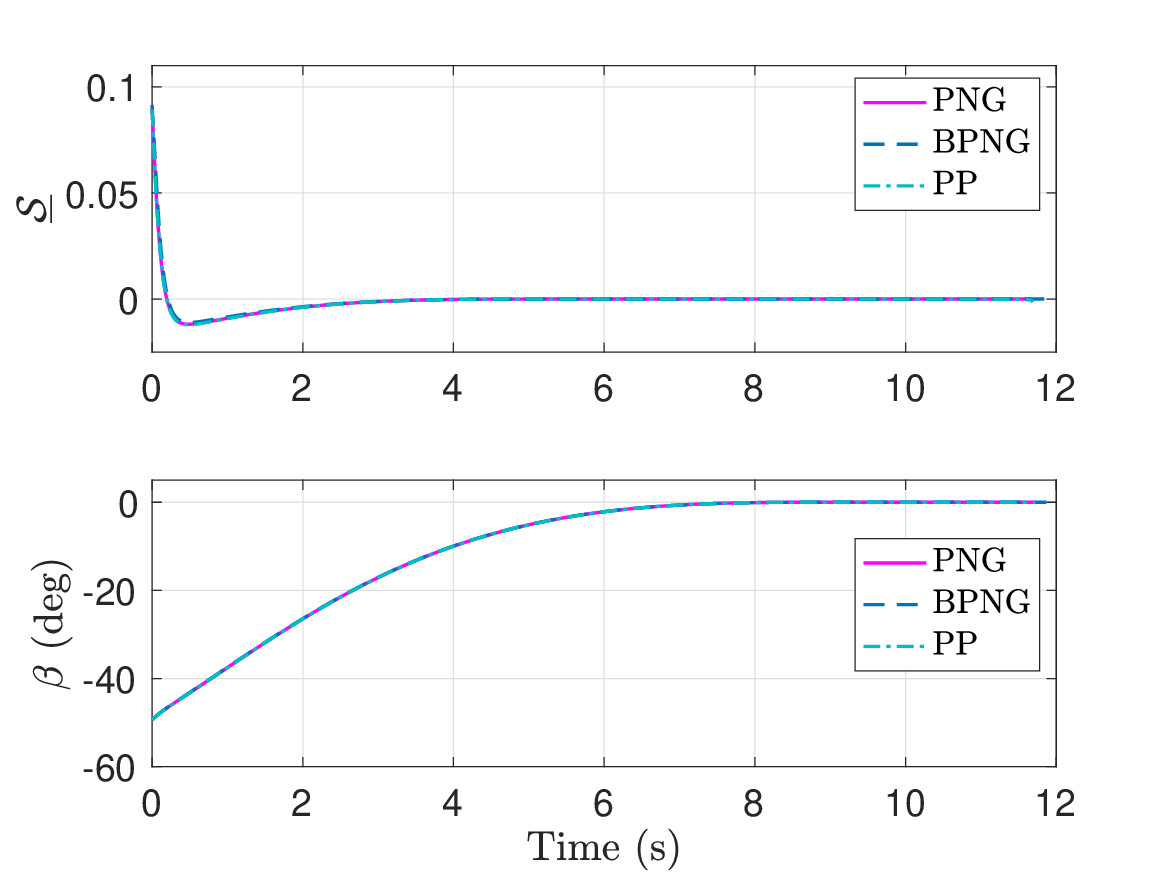}
    \caption{Sliding manifold and error.}
    \label{fig:p_diff_guid_s_beta}
    \end{subfigure}%
    \begin{subfigure}{.25\linewidth}
    \centering
    \includegraphics[width=\linewidth]{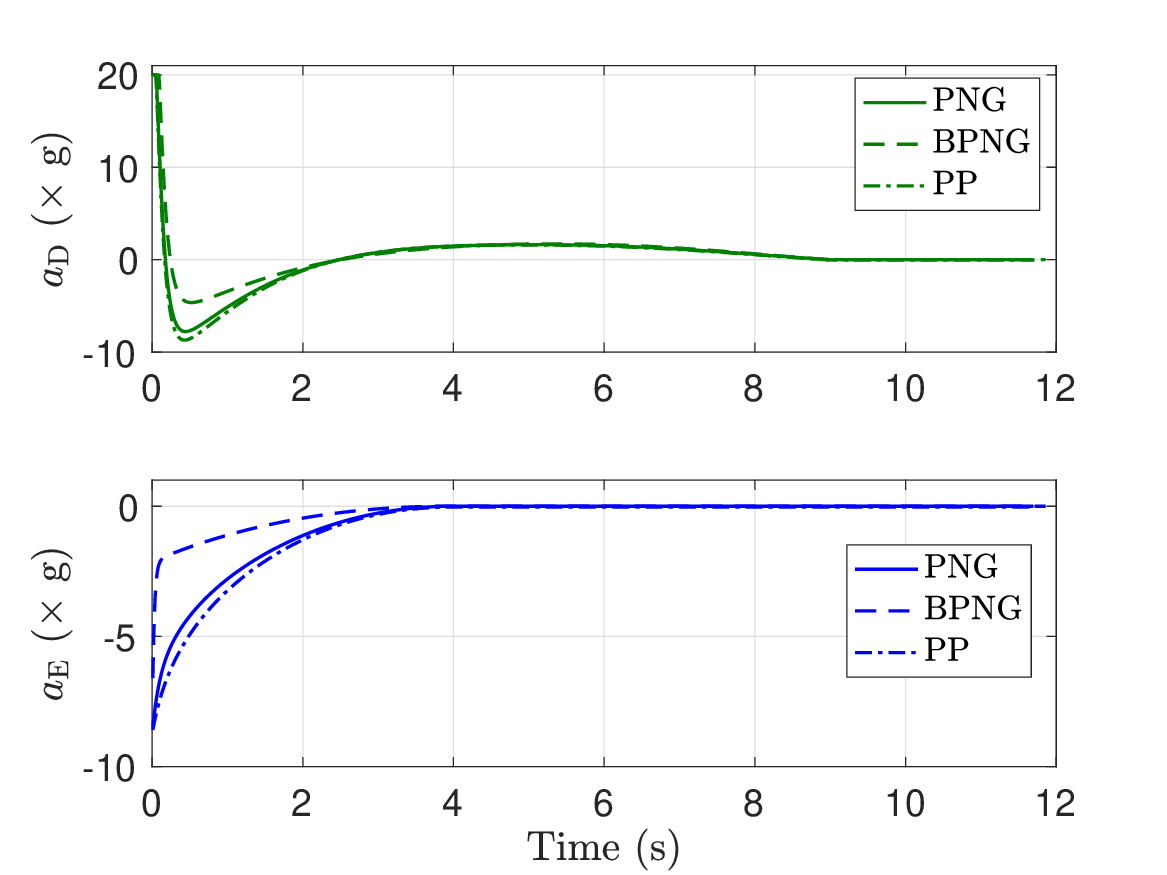}
    \caption{Steering controls $a_{\mathrm{D}}$ and $a_{\mathrm{E}}$.}
    \label{fig:p_diff_guid_ad_ae}
    \end{subfigure}%
    \begin{subfigure}{.25\linewidth}
    \centering
    \includegraphics[width=\linewidth]{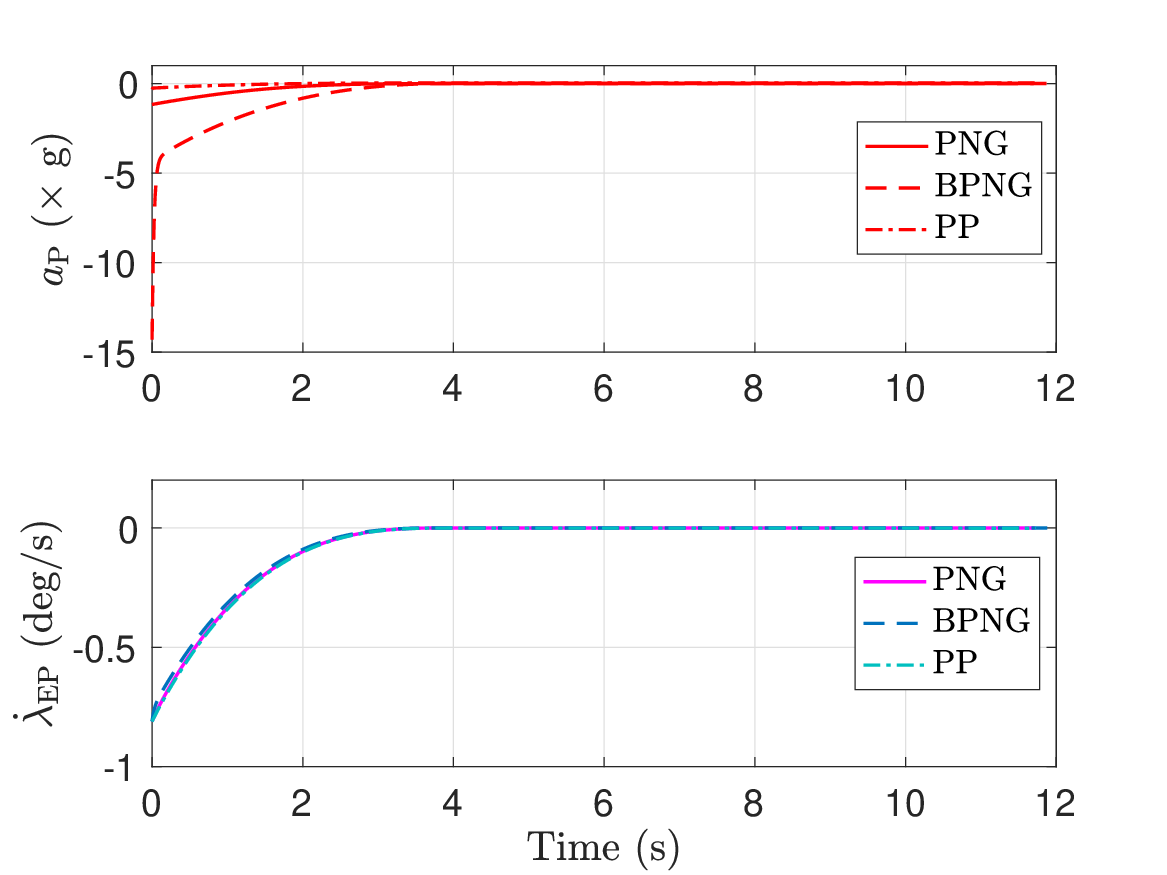}
    \caption{$a_{\mathrm{P}}$ and $\dot{\lambda}_{\mathrm{EP}}$.}
    \label{fig:p_diff_guid_ap_dotlambda}
    \end{subfigure}
\caption{The pursuer is using different guidance strategy to capture the evader.}
\label{fig:p_diff_guidance}
\end{figure*}

\subsection{Information-level Cooperation between the Evader and Defender}
In this section, we demonstrate the effectiveness of the proposed guidance strategy when the evader and defender employ information-level cooperation to capture the pursuer. The design parameters are chosen as: $\mathcal{K}_{1}=1$, $t_2=4\,\si{S}$, $k_3=3$, $t_1=5\,\si{s}$, $t^{\star}=9\,\si{s}$, $k_1=k_2=3$, and $\mathcal{K}=1$.

Under the guidance commands derived in \Cref{thm:aD,thm:aE}, the performance of the information-level cooperation strategy is evaluated for different values of the desired angle $\chi^{\star}$ while maintaining the same initial engagement geometry. The values of $\chi^{\star}$ are chosen to be $150^\circ$, $180^\circ$, and $200^\circ$, respectively and the performance is depicted in \Cref{fig:p_diff_chi}. In this engagement scenario, the evader is initially positioned at the origin with a heading angle of $0^\circ$. The defender is placed at a radial distance of $500\,\si{m}$ from the evader with a LOS angle of $45^\circ$ and a heading of $45^\circ$, while the pursuer starts at a radial distance of $5000\,\si{m}$ from the evader with a heading of $180^\circ$, that is, directly pointing towards the evader. It can be observed from \Cref{fig:p_diff_chi_trajectory} that the defender successfully captures the pursuer for all values of $\chi^{\star}$. As shown in \Cref{fig:p_diff_chi_ad_ae,fig:p_diff_chi_ap_dotlambda}, the evader nullifies its LOS rate with respect to the pursuer exactly at $t_2 = 4\,\si{s}$ by applying the guidance command given in \eqref{eq:ae_passive}. In other words, within $4\,\si{s}$, the evader becomes non-maneuvering relative to the pursuer, effectively luring it by acting as bait. This behavior is further evident in \Cref{fig:p_diff_chi_ad_ae}, where the evader’s guidance command reduces to zero precisely at $t = 4\,\si{s}$ and remains there. During the same period of time, the pursuer applies a non-zero control input in an attempt to capture the evader. However, once the evader regulates its LOS rate relative to the pursuer, the situation becomes favorable for the pursuer. Believing interception to be assured, the pursuer also ceases maneuvering and maintains zero control input from $t = 4\,\si{s}$ onward, as shown in \Cref{fig:p_diff_chi_ap_dotlambda}. Meanwhile, \Cref{fig:p_diff_chi_s_beta} illustrates that the sliding mode is enforced on the inner-layer sliding manifold at $t_{1} = 5\,\si{s}$, and the error $\beta$ converges to zero exactly at $t^{\star} = 9\,\si{s}$ for all values of the desired angle $\chi^\star$. Consequently, the defender aligns itself with the desired course at the same instant, $t^{\star} = 9\,\si{s}$, which ultimately leads to the successful capture of the pursuer for all three values of $\chi^\star$.

We further demonstrate the performance of the proposed ({information-level} cooperation) guidance strategy when the evader positions itself as bait at different time instants $t_{2}$. The values of $t_{2}$ are chosen as $4\,\si{s}$, $6\,\si{s}$, and $8\,\si{s}$, respectively. Keeping the other engagement parameters identical to the previous case, the results are illustrated in \Cref{fig:p_diff_t2}. It can be observed from \Cref{fig:p_diff_t2_trajectory} that the defender successfully captures the pursuer for all choices of $t_2$. Specifically, the evader nullifies its LOS rate with the pursuer exactly at the prescribed time instant $t_2$ by applying the guidance command in \eqref{eq:ae_passive}. For smaller values of $t_2$, the evader requires a larger initial control effort. However, once the LOS rate is nullified, the demand reduces to zero. In response, the pursuer initially applies a higher value guidance command, but as soon as the evader becomes non-maneuvering (precisely at $t_2$), it transitions to a straight-line path to capture the evader. By contrast, the defender utilizes the maneuvering information of the evader and always captures the pursuer, exhibiting a similar behavior across all cases. 

Finally, we evaluate the performance of the proposed strategy when the pursuer employs different guidance laws to capture the evader. For simplicity, the engagement geometry is kept identical to the previous case. Here, we consider that the pursuer is using the three different strategies: PNG, Biased PNG, and PPG. In the case of BPNG, it is assumed that the pursuer has access to the evader’s maneuver information and incorporates a bias term into PNG to anticipate and counter the evader’s maneuvers. Under the proposed information level cooperation strategy, performance of the proposed strategy for such engagement scenarios is shown in \Cref{fig:p_diff_guidance}. As observed from \Cref{fig:p_diff_guidance}, the defender successfully intercepts the pursuer at the desired angle $\chi^\star=180^\circ$ in all cases, despite the different guidance strategies adopted by the pursuer. One may notice from \Cref{fig:p_diff_guid_ap_dotlambda} that during the initial period, the pursuer applies different values of the guidance command to capture the evader. However, once the pursuer lures it onto the collision course (by nullifying its LOS rate with the pursuer to zero) at $t_2=4\,\si{s}$, the pursuer also stops maneuvering and aims to capture the evader within the shortest possible time by moving in a straight line directly towards the evader. Nonetheless, the defender, leveraging the information provided by the evader, intercepts the pursuer in all scenarios.

It is important to note that the defender achieves interception with only information-level cooperation from the evader, irrespective of the guidance strategy employed by the pursuer. This, in turn, bolsters the claim that the proposed guidance strategies are independent of the pursuer’s strategy.
\section{Conclusions}\label{sec:conclusion}
In this paper, we have introduced a geometric approach that addresses the challenge of pursuit-evasion scenarios involving three agents with arbitrary initial geometries. Our proposed solution, the evader-defender cooperative guidance strategy, offers an effective means to guarantee pursuit-evasion under diverse conditions. Specifically, it guarantees that the defender reaches a specific angle precisely within a predefined time, irrespective of the initial engagement geometry, thus intercepting the pursuer before it can capture the evader. A distinguishing feature of our approach is its adaptability to scenarios characterized by nonlinear dynamics, non-holonomic constraints, and large heading angle errors, which are often encountered in practical motion control situations, including aircraft defense. This paper presents a robust and practical solution for guaranteed pursuit-evasion problems, and also lays the foundation for identifying crucial conditions and configurations that facilitate successful evasion or capture. Analyzing geometrical solutions in three-dimensional settings could be an interesting future research topic.
\bibliographystyle{IEEEtaes}
\bibliography{pursuit_evasion.bib}

\begin{thebibliography}{10}
\providecommand{\url}[1]{#1}
\csname url@samestyle\endcsname
\renewcommand{\newblock}{\par}
\providecommand{\bibinfo}[2]{#2}
\providecommand{\BIBentrySTDinterwordspacing}{\spaceskip=0pt\relax}
\providecommand{\BIBentryALTinterwordstretchfactor}{4}
\providecommand{\BIBentryALTinterwordspacing}{\spaceskip=\fontdimen2\font plus
\BIBentryALTinterwordstretchfactor\fontdimen3\font minus
  \fontdimen4\font\relax}
\providecommand{\BIBforeignlanguage}[2]{{%
\expandafter\ifx\csname l@#1\endcsname\relax
\typeout{** WARNING: IEEEtran.bst: No hyphenation pattern has been}%
\typeout{** loaded for the language `#1'. Using the pattern for}%
\typeout{** the default language instead.}%
\else
\language=\csname l@#1\endcsname
\fi
#2}}
\providecommand{\BIBdecl}{\relax}
\BIBdecl

\bibitem{isaacs1999differential}
R.~Isaacs
\newblock \emph{Differential games: a mathematical theory with applications to
  warfare and pursuit, control and optimization}. Courier Corporation, 1999.

\bibitem{7855582}
S.~Y. Hayoun, M.~Weiss, and T.~Shima
\newblock  A mixed ${L}_2$/${L}_\alpha$ differential game approach to
  pursuit-evasion guidance \newblock  \emph{IEEE Transactions on Aerospace and
  Electronic Systems}, vol.~52, no.~6, pp. 2775--2788, 2016.

\bibitem{4383577}
T.~Shima and O.~M. Golan
\newblock  Linear quadratic differential games guidance law for dual controlled
  missiles \newblock  \emph{IEEE Transactions on Aerospace and Electronic
  Systems}, vol.~43, no.~3, pp. 834--842, 2007.

\bibitem{4101686}
R.~L. Boyell
\newblock  Defending a moving target against missile or torpedo attack
  \newblock  \emph{IEEE Transactions on Aerospace and Electronic Systems},
  vol.~12, no.~4, pp. 522--526, 1976.

\bibitem{4102335}
R.~L. Boyell
\newblock  Counterweapon aiming for defense of a moving target \newblock
  \emph{IEEE Transactions on Aerospace and Electronic Systems}, vol.~16, no.~3,
  pp. 402--408, 1980.

\bibitem{doi:10.2514/1.G001083}
E.~Garcia, D.~W. Casbeer, and M.~Pachter
\newblock  Cooperative strategies for optimal aircraft defense from an
  attacking missile \newblock  \emph{Journal of Guidance, Control, and
  Dynamics}, vol.~38, no.~8, pp. 1510--1520, 2015.

\bibitem{doi:10.2514/1.51765}
T.~Shima
\newblock  Optimal cooperative pursuit and evasion strategies against a homing
  missile \newblock  \emph{Journal of Guidance, Control, and Dynamics},
  vol.~34, no.~2, pp. 414--425, 2011.

\bibitem{doi:10.2514/1.49515}
V.~Shaferman and T.~Shima
\newblock  Cooperative multiple-model adaptive guidance for an aircraft
  defending missile \newblock  \emph{Journal of Guidance, Control, and
  Dynamics}, vol.~33, no.~6, pp. 1801--1813, 2010.

\bibitem{doi:10.2514/1.58531}
O.~Prokopov and T.~Shima
\newblock  Linear quadratic optimal cooperative strategies for active aircraft
  protection \newblock  \emph{Journal of Guidance, Control, and Dynamics},
  vol.~36, no.~3, pp. 753--764, 2013.

\bibitem{doi:10.2514/1.61832}
S.~Rubinsky and S.~Gutman
\newblock  Three-player pursuit and evasion conflict \newblock  \emph{Journal
  of Guidance, Control, and Dynamics}, vol.~37, no.~1, pp. 98--110, 2014.

\bibitem{6315051}
T.~Yamasaki and S.~N. Balakrishnan
\newblock  Terminal intercept guidance and autopilot for aircraft defense
  against an attacking missile via {3D} sliding mode approach \newblock  In
  \emph{American Control Conference (ACC)}, 2012, pp. 4631--4636.

\bibitem{doi:10.2514/1.G000659}
S.~R. Kumar and T.~Shima
\newblock  Cooperative nonlinear guidance strategies for aircraft defense
  \newblock  \emph{Journal of Guidance, Control, and Dynamics}, vol.~40, no.~1,
  pp. 124--138, 2017.

\bibitem{9274339}
A.~Sinha, S.~R. Kumar, and D.~Mukherjee
\newblock  Cooperative salvo based active aircraft defense using impac time
  guidance \newblock  \emph{IEEE Control Systems Letters}, vol.~5, no.~5, pp.
  1573--1578, 2021.

\bibitem{doi:10.1007/s10846-022-01570-y}
A.~Sinha, S.~R. Kumar, and D.~Mukherjee
\newblock  Three-agent time-constrained cooperative pursuit-evasion \newblock
  \emph{Journal of Intelligent \& Robotic Systems}, vol. 104, no.~2, p.~28,
  2022.

\bibitem{doi:10.1016/j.ast.2020.105787}
X.~Yan and S.~Lyu
\newblock  A two-side cooperative interception guidance law for active air
  defense with a relative time-to-go deviation \newblock  \emph{Aerospace
  Science and Technology}, vol. 100, p. 105787, 2020.

\bibitem{doi:10.2514/6.2010-7876}
T.~Yamasaki and S.~Balakrishnan
\newblock  Triangle intercept guidance for aerial defense \newblock  In
  \emph{AIAA Guidance, Navigation, and Control Conference}, 2010, p. 7876.

\bibitem{9301417}
S.~R. Kumar and D.~Mukherjee
\newblock  Cooperative active aircraft protection guidance using line-of-sight
  approach \newblock  \emph{IEEE Transactions on Aerospace and Electronic
  Systems}, vol.~57, no.~2, pp. 957--967, 2021.

\bibitem{doi:10.2514/1.58566}
T.~Yamasaki, S.~N. Balakrishnan, and H.~Takano
\newblock  Modified command to line-of-sight intercept guidance for aircraft
  defense \newblock  \emph{Journal of Guidance, Control, and Dynamics},
  vol.~36, no.~3, pp. 898--902, 2013.

\bibitem{10643724}
X.~Wang, M.~Yang, S.~Wang, and T.~Chao
\newblock  Two-stage game guidance strategy with impact point for active
  defense aircraft in two-on-two engagement \newblock  \emph{IEEE Transactions
  on Aerospace and Electronic Systems}, vol.~61, no.~1, pp. 710--729, 2025.

\bibitem{doi:10.2514/1.G006705}
S.~R. Kumar and D.~Mukherjee
\newblock  Generalized triangle guidance for safeguarding target using barrier
  {L}yapunov function \newblock  \emph{Journal of Guidance, Control, and
  Dynamics}, vol.~45, no.~11, pp. 2193--2201, 2022.

\bibitem{9122473}
E.~Garcia, D.~W. Casbeer, A.~Von~Moll, and M.~Pachter
\newblock  Multiple pursuer multiple evader differential games \newblock
  \emph{IEEE Transactions on Automatic Control}, vol.~66, no.~5, pp.
  2345--2350, 2021.

\bibitem{doi:10.1016/j.automatica.2011.06.010}
S.~D. Bopardikar, S.~L. Smith, and F.~Bullo
\newblock  On vehicle placement to intercept moving targets \newblock
  \emph{Automatica}, vol.~47, no.~9, pp. 2067--2074, 2011.

\end{thebibliography}

\end{document}